\documentclass{article}
\usepackage{graphicx}
\usepackage{graphicx}
\usepackage{float}
\usepackage{amsfonts,amsmath,amssymb}
\usepackage[T1]{fontenc}
\usepackage{lmodern}
\usepackage{overpic}
\usepackage{pstricks}
\usepackage{mathtools}
\usepackage{epsfig}
\date{\today}

\newcommand{\insertplot}[5]{\begin{figure}
 \hfill\hbox to 0.05in{\vbox to #5in{\vfill
 \inputplot{#1}{#4}{#5}}\hfill}
 \hfill\vspace{-.1in}
 \caption{#2}\label{#3}
 \end{figure}}
 \newcommand{\inputplot}[3]{
 \special{ps: plotfile #1}
}
\newcounter{fig}

\newcommand{\diff}{\mathrm{d}}
\newcommand{\be}{\begin{equation}}
\newcommand{\ee}{\end{equation}}
\newcommand{\bea}{\begin{eqnarray}}
\newcommand{\eea}{\end{eqnarray}}

\begin{document}
 \title{ 
Nutty black holes in galileon scalar-tensor gravity 
} 
\author{
{\large A. Brandelet, Y. Brihaye}$^{\dagger}$, T. Delsate 
{\large } 
and {\large L.Ducobu}$^{\dagger}$  
\\ 
\\
$^{\dagger}${\small Physique-Math\'ematique, Universit\'e de
Mons, Mons, Belgium}
}
\maketitle
 
\begin{abstract} 
Einstein gravity supplemented by a scalar field non-minimally  coupled to a Gauss-Bonnet term
provides an example of model of scalar-tensor gravity where hairy black holes do exist. We consider
the classical equations within a metric endowed with a NUT-charge and obtain a two-parameter family of
nutty-hairy black holes. The pattern  of these solutions in the exterior and the interior of their horizon is studied
in some details. The influence of both -- the hairs and the NUT-charge -- on the lightlike and timelike geodesics is emphasized.  
\end{abstract}

\section{Introduction}
Evading the ``No-Hair-Theorem'' for black holes in General Relativity -- 
and its numerous extended versions~-- has constituted a  challenge for a long time.
One issue consists in supplementing gravity by an appropriate matter sector like the Skyrme
Lagrangian \cite{moss}. Recently several kinds of hairy black holes have been constructed with
a simpler matter sector~: scalar fields.

Both cosmological and astrophysical observations suggest the presence
of scalar fields in the models attempting to describe the Universe in its early (inflaton, dilaton)
or actual stage (dark matter). 
These scalar fields could be fundamental (although not yet directly observed) or effective, modelling
the effects of  more involved -- but still unknown -- phenomena on space-time and standard particles. 
These considerations, namely, motivate the extension of standard  formulation of General Relativity
(also called tensor gravity) to the most general scalar-tensor gravity theory.

The first general construction in this direction was achieved by G. Horndeski  in \cite{horndeski}
where the condition of second order equations is imposed throughout. 
Recently, new families of scalar-tensor theories, the so-called Galileon \cite{Nicolis:2008in} 
and generalized Galileon \cite{Deffayet:2011gz},
have been proposed with different motivations and contexts (for a review see e.g. \cite{Deffayet:2013lga}). 
In particular, these theories require
a symmetry of the Lagrangian under the shift $\phi \rightarrow \phi + C$ where $\phi$ denotes
the scalar field and $C$ a constant. 
In four dimensions \cite{kobayashi}, the generalized Galileon theory has been shown to be equivalent to the Horndeski theory. 
The generalized Galileon theory is quite general, involving the different geometric invariants and
depending on several arbitrary functions of the standard kinetic
terms $\partial_{\mu} \phi \partial^{\mu} \phi $.
A no-hair theorem for black holes in generic forms of the Galileon theory, assuming  static spherically symmetric space-time and scalar field, was established  \cite{Hui:2012qt}.
However, as shown in  \cite{Sotiriou:2013qea,Sotiriou:2014pfa} a few specific choices of the Galileon Lagrangian
allow for hairy black holes to exist.

The hairy black holes constructed in the framework of Galileon gravity have a real and massless scalar field
and can be static. Independently of the shift symmetry of the scalar field  hypothesis, another class of models
that retained a lot of attention is the Einstein-Gauss-Bonnet-Dilaton theory where hairy black holes
can  be constructed as well \cite{Kleihaus:2011tg,Blazquez-Salcedo:2016yka}.
 Hairy black holes have been constructed within Einstein gravity coupled to a complex scalar field
 in \cite{Herdeiro:2014goa}
 (see also \cite{Herdeiro:2015waa} for a review). In this case, the scalar field needs to have a  mass and
 the black hole exists only when it rotates quickly enough.  

In the long history of classical solutions of General Relativity, the so-called ``NUT solution'' \cite{NUT}
is certainly one of the most intriguing. In the absence of matter fields, the NUT solution is a generalization
of the Schwarzschild black hole characterized by a new parameter~: the so-called  NUT charge  $n$.
Although purely analytic, the NUT space-time presents peculiarities \cite{Misner,misner-book} that makes that its physical interpretation is, till now, a matter of debate. In particular, the solution presents a Misner string singularity on the polar axis
and the corresponding space-time contains closed timelike curves. 
Various arguments rehabilitating space-time with a NUT charge are proposed in \cite{Clement:2015cxa}.
In spite of the difficulty of finding a global definition of the NUT space-time,
the solution possesses many remarquable properties, namely~:
(i) like the Kerr solution, it is stationary but non-static due to non-vanishing $g_{t \varphi}$ metric terms~;
(ii) it can be extended analytically 
(i.e. without curvature singularity)  in the interior region  by means of a TAUB solution.

Likely for these reasons,  several authors
(see namely \cite{Kagramanova:2010bk, Jefremov:2016dpi}) have considered the NUT parameter
as a possible ingredient of some astrophysical object and have studied its effect on 
geodesics  in NUT space-times. 
Another application of the NUT parameter was proposed recently in \cite{Brihaye:2016lsx} to obtain
families of 
 non-trivial, spherically symmetric solutions of the Einstein-Chern-Simons gravity
 coupled to a scalar field. Such a construction was possible by
taking advantage of the stationary character of the underlying metric. 

In this paper we extend the construction of the hairy black holes of \cite{Sotiriou:2014pfa} by including  a NUT parameter in
the  metric. We show that Nutty-hairy-black holes exist in a specific domain of the NUT charge and Gauss-Bonnet parameter.  
A special emphasis is set on the way the NUT charge affects the solution in the interior of the black hole. 
Also, we study the influence on the light-like geodesic of both the presence of the scalar field and of the NUT charge. 
It is found in particular that, mimicking a rotation, the NUT charge leads to a non-planar drift of the trajectories. 

The paper is organized as follows. In Sect.~2 we present the model, the ansatz for the metric, the boundary conditions of the ensuing classical equations and sketch the form of a perturbative solution.
The non-perturbative solutions, obtained with a numerical method, are reported in Sect.~3. 
The influence of the Gauss-Bonnet gravity term and of the NUT parameter on the light-like geodesics 
are emphasized in Sect.~4 and illustrated by some figures.
Conclusion and perspectives are given in Sect.~5. 
  
\section{The model}
\label{sectmodel}
\subsection{The Gauss-Bonnet modified gravity}
 The modified theory that we want to emphasize was first studied in \cite{Sotiriou:2014pfa}
 as a particular case of the general scalar-tensor-galileon-gravity. It can be defined in 
terms of its action:
\be
\label{CSaction}
S \coloneqq S_{\rm EH} + S_{\rm GB} +  S_{\phi} ,
\ee
where the Einstein-Hilbert term is given by
\be
\label{EH-action}
S_{\rm{EH}} = \kappa \int_{{\cal{V}}} \diff^4x  \sqrt{-g}  R, \ \ 
\ee
the Gauss-Bonnet term is given by
\be
\label{gb-action}
S_{\rm{GB}} =  \frac{\gamma}{2} \int_{{\cal{V}}} \diff^4x  \sqrt{-g} \phi {\cal G} \ \ , \ \ 
{\cal G} = (R^{\mu \nu \rho \sigma} R_{\mu \nu \rho \sigma} - 4 R^{\mu \nu}R_{\mu \nu} + R^2 ),
\ee
the scalar field term is given by
\be 
\label{Theta-action}
S_{\phi} = - \beta \frac{1}{2} \int_{{\cal{V}}} \diff^{4}x \sqrt{-g} \left[ g^{\rho\sigma}
\left(\nabla_{\rho} \phi \right) \left(\nabla_{\sigma} \phi \right)  \right].
\ee
 In these equations, $\kappa^{-1} = 16 \pi G$, $\gamma$ and $\beta$ are dimensional coupling constants, $g$ is the determinant of the metric, $\nabla_{\mu}$ is the covariant derivative associated with $g_{\mu\nu}$, $R$ is the Ricci scalar
  and the volume integrals are on the manifold ${{\cal{V}}}$.

The scalar field $\phi$  is in principle a function of space-time.
In the case $\phi = \textrm{const.}$, the Gauss-Bonnet modified gravity 
would reduce identically to GR since, as well known (see e.g. Appendix B in \cite{Sotiriou:2014pfa}), 
the Gauss-Bonnet action density  
(\ref{gb-action})
 can be expressed as a divergence.
 
The equations of motion for this model read
\be
\label{eq:eom}
 G_{\mu\nu} + \frac{\gamma}{4 \kappa} K_{\mu\nu} = \frac{\beta}{2\kappa}  T_{\mu\nu},
\ee
where $G_{\mu\nu}=R_{\mu\nu}-\frac{1}{2} g_{\mu\nu}R$ is the Einstein tensor, the tensor
\be
       K_{\mu\nu} = (g_{\mu \lambda} g_{\nu \delta} + g_{\mu \delta} g_{\nu \lambda})\nabla_{\rho} (\partial_\sigma \phi \epsilon^{\sigma \delta \xi \chi} \epsilon^{\lambda \rho \omega \eta} R_{\omega \eta \xi \chi}), 
\ee
where $\epsilon^{\alpha \beta \mu \nu}$ is the Levi-Civita tensor, results from the variation of the Gauss-Bonnet term and
$T_{\mu\nu} $ is the energy-momentum tensor of the scalar field : 
\be
\label{Tab-theta}
T_{\mu\nu} 
=   \left[  \left(\nabla_{\mu} \phi\right) \left(\nabla_{\nu} \phi\right) 
    - \frac{1}{2}  g_{\mu\nu}\left(\nabla_{\rho} \phi\right) \left(\nabla^{\rho} \phi\right) 
  \right].
\ee

The vanishing of the variation of the action also leads to an extra equation of motion for the
Gauss-Bonnet coupling field, namely 
\be 
\label{eq:constraint}
\beta \; \square \phi = - \frac{\gamma}{2} {\cal G},
\ee
where $\square = \nabla^\mu \nabla_\mu$, which we recognize as the Klein-Gordon equation in the presence of  a sourcing term. 

{\subsection{The metric}}
We consider NUT-charged space-times \cite{NUT,Misner} whose metric can be written locally in the
form
\begin{eqnarray}
\label{metric} 
ds^2=\frac{dr^2}{N(r)}+P^2(r)(d\theta^{2}+\sin
^2\theta d\varphi^{2}) -N(r)A^2(r)(dt+4 n
\sin^2(\frac{\theta}{2}) d\varphi)^{2},
\end{eqnarray}
the NUT parameter $n$ being defined as usual in terms of the coefficient
appearing in the differential $dt+4 n \sin^2(\theta/2)d\varphi$.
Here $\theta$ and $\varphi$ are the standard angles parametrizing 
an $S^2$ sphere with ranges $0 \leq \theta \leq \pi$, $0 \leq \varphi \leq 2\pi$.
Apart from the Killing vector $K_0=\partial_{t}$, this line element possesses
three more Killing vectors
characterizing the NUT symmetries :
\begin{eqnarray}
\label{Killing}
\nonumber
K_1&=&\sin \varphi \partial_{\theta}
+\cos \varphi \cot \theta  \partial_{\varphi}
+2n\cos\varphi \tan \frac{\theta}{2}  \partial_{t},
\\
K_2&=&\cos \varphi \partial_{\theta}
-\sin \varphi \cot \theta\partial_{\varphi}
-2n\sin \varphi\tan \frac{\theta}{2}\partial_{t},
\\
\nonumber
K_3&=&\partial_{\varphi}-2n\partial_{t}.
\end{eqnarray}
Unexpectedly, these  Killing vectors form a subgroup with the same structure constants
that are obeyed by spherically symmetric
solutions $[K_i,K_j]=\varepsilon_{ijk}K_k$.

The $n\sin^2 (\theta/2)$ term in the metric means that a small loop around the
$z-$axis does not shrink to zero at $\theta=\pi$.
This singularity can be regarded as the analogue of a Dirac string in electrodynamics
 and is not related to the usual degeneracies of spherical coordinates on the two-sphere.
This problem was first encountered in the vacuum NUT metric.
One way to deal with this singularity has been proposed by Misner \cite{misner-book}.
His argument holds also independently of the precise functional form of $N$ and $A$.
In this construction, one considers one coordinate patch in which the string runs off to
infinity along the north axis.
A new coordinate system can then be
found with the string running off to infinity along the south axis
with $t'=t+4n\varphi,$ the string becoming an artifact resulting 
from a poor choice of coordinates.
It is clear that the 
$t$ coordinate is also periodic with period
$8 \pi n$ and essentially becomes an Euler angle coordinate on $S^3$.
Thus an observer with $(r,\theta,\varphi)=const.$ follows a closed timelike curve.
These lines cannot be removed by going to a covering space
and there is no reasonable spacelike surface.
One finds also that surfaces of constant radius have the topology
of a three-sphere, in which there is a Hopf fibration of the $S^1$
of time over the spatial $S^2$ \cite{misner-book}.

Therefore for $n$ different from zero, the  metric structure
(\ref{metric}) generically shares the same troubles exhibited by the vacuum
Taub-NUT gravitational field \cite{Mueller:1985ij}, and the solutions cannot be
interpreted properly as black holes.

The vacuum Taub-NUT one corresponds to
\begin{eqnarray}
\label{vacuum}
A(r)=1~~~,~~~\phi(r)=0~~~, ~~~ P(r)^2 = n^2+r^2~~~,~~~
N(r)=1-\frac{2(Mr+n^2)}{r^2+n^2}.
\end{eqnarray}
where $M = (r_h^2 - n^2)/(2r_h)$. This solution presents an horizon at $r=r_h$. 
In the following section, we will study how these closed form solutions get 
deformed by the Gauss-Bonnet term.

{\subsection{Gauge fixing and boundary conditions}\label{sectbound}}
Up to our knowledge, the system above does not admit closed form solutions for $\gamma > 0$.
The solutions can then be constructed either perturbatively (for instance using $\gamma$ as a perturbative parameter)
either non perturbatively by solving the underlying boundary-value-differential equations numerically. 

For the numerical integration, the ``gauge'' freedom associated with  the redefinition of the radial coordinate $r$ has to be fixed. 
We found it convenient to fix  this freedom by setting $P(x)^2 = x^2 + n^2$ and to  note $x$
the radial coordinate defined this way. 
The ansatz is then completed by assuming the scalar field of the form $\phi(x^{\mu}) = \phi(x)$. 

The Einstein-Gauss-Bonnet-Klein-Gordon equations can then be transformed into a system of 
three coupled  differential equations for the functions $N(x)$, $A(x)$ and $\phi(x)$. 
The equations for the metric functions are of the first order while the Klein-Gordon equation is, 
as usual, of the second order.

Our  goal is to construct regular solutions presenting (like black holes)
a horizon at $x = x_h$, the regularity of the solution at the horizon
 needs the following condition  to be 
imposed~:
\be
\label{regularity}
      N(x_h)=0 \ \ ,  \ \ 
     \left[ \gamma (\phi')^2 + x \phi ' + 3 \gamma \frac{x^2 - n^2 A^2}{(x^2 + n^2)^2} \right]_{x=x_h} = 0 \ \ , \ \ \phi(x_h)=0 \ ,
\ee
(the last relation is imposed by using the  invariance of the theory under the translations
of the scalar field);
these three relations are completed by a fourth condition at infinity, namely $A(\infty) = 1$. 

Also, the equations are invariant under the following rescaling by the parameter $\lambda$ :
\be
\label{rescaling}
  x \rightarrow \lambda x \ \ , \ \ n \rightarrow \lambda n \ \ , \ \ P \rightarrow \lambda^2 P \ \ , \ \ 
  \gamma \rightarrow \lambda^2 \gamma .
\ee
This symmetry can be exploited to fix one of the three parameters  $x_h$, $n, \gamma$ to a particular
value, reducing by one unit the number of parameters to vary. We will use it by setting $x_h=1$ throughout
the rest of the paper.

In the asymptotic region, the fields obey  the following form~:
\begin{eqnarray}
    N(x) &=& 1 - \frac{2M}{x} + \frac{Q^2- 4 n^2}{2 x^2} + \frac{M(Q^2+4n^2)}{2 x^3} + o(1/x^4)  \ \ , \ \ \nonumber \\
    A(x) &=& 1 - \frac{Q^2}{4 x^2} - \frac{2 M Q^2}{3 x^3} + o(1/x^4)  \ \ , \ \ \nonumber \\
    \phi(x) &=& \phi_{\infty} + \frac{Q}{x} + \frac{QM}{x^2} + \frac{Q(4n^2 + 16 M^2 - Q^2)}{12 x^3} +   o(1/x^4) ~.
    \label{asymptotic}
\end{eqnarray}
The perturbative expansion depends on the two ``charges'' $M$ and $Q$ which are determined numerically.
\subsection{Invariants}
For later use, we mention  the Ricci and Kretchmann invariants for the metric (\ref{metric})
\be
\label{ricci}
R = -\frac{P^3 \left(2 N P A''+A' \left(3 P N'+4 N P'\right)\right)-2 n^2 A^3 N+A P^2 \left(P^2 N''+4 P \left(N' P'+N P''\right)+2 N
 P'^2-2\right)}{A P^4},
\ee
where the prime denotes the derivative with respect to $x$.
The expression for the Kretschmann invariant is much longer and we do not write it.

In the case of the vacuum NUT solution (\ref{vacuum})
the Ricci scalar is identically zero while for the Kretschmann invariant we find
\be
\label{kret}
K = \frac{K_{num}}{x_h^2(n^2+x^2)^6}~,
\ee
with
\be
K_{num} = 12 n^6(n^4 - 6 n^2 x_h^2 + x_h^4)(x^6 -15n^2 x^4 + 15 n^5 x^2 - n^6) 
+ 96 x_h n^4 (x_h^2 - n^2) (3 x^5 - 10 n^2 x^3 + 3 n^4 x) .
\ee

{\subsection{Perturbative expansion.}}
The Einstein-Gauss-Bonnet equations can be attempted to be solved perturbatively in powers of 
the Gauss-Bonnet coupling constant $\gamma$. The perturbation has the same form as in the $n=0$ case
\cite{Sotiriou:2014pfa}~:
\be
     N(x) = 1-\frac{2(Mx+n^2)}{x^2+n^2} + \gamma^2 (\Delta N)_1 + o(\gamma^4) \ \ , \ \ 
     A(x) = 1 + \gamma^2 (\Delta A)_1 + o(\gamma^4) \ \ , \ \ \phi(x) = \gamma \phi_1(x) + o(\gamma^3) \ .
\label{perturbation}
\ee
Already at the first order, the form of the scalar field is rather involved although easy to construct~:
\be
    \phi'_1 = \frac{P(x,n)}{(n^2+x^2)^4(n^2+x)}~,
\ee
with
\be
    P(x,n) = - x^7 - x^6 - x^5(1+4n^2)-x^4 n^2 (10-n^2) + x^3 n^2(8 - 23 n^2) + x^2 n^4(25-8n^2) - x n^4(3-16n^2)+ n^6(3n^2-2) .  \ee
For simplicity we have written the derivative $\phi'$ which is the function entering effectively in  the Lagrangian.
The integration constant was fixed in such a way that the field is regular at $x=x_h$.
 The above expression is regular at $x=0$ for $n>0$ and coincides with \cite{Sotiriou:2014pfa} in the limit $n\to 0$.
 Confirming the non-pertubative results discussed in the next section,
  the Taub-Nut parameter $n$ regularizes the scalar field at the origin.
The form of $\Delta N$ and $\Delta A$ is much more involved and will not be reported here.
However, for a crosscheck of the numerical results, we mention that these functions are singular
in the limit $x \to 0$. For instance, we find $\Delta A = \Gamma/x^2 + o(1/x)$ where $\Gamma$ is a function of $n$.
{\section{Numerical Results}\label{sectnum}}
The system of differential equations above cannot be solved explicitly for generic value of
the three external parameters $x_h, \gamma, n$. We therefore used a numerical routine to construct
the solutions. The subroutine COLSYS \cite{COLSYS} based on collocation method 
with a self-adapting mesh has been used for the computation. 
 
The integration proceeds in two steps. First, we
solve the equations for  $x \in [x_h,\infty]$
with a particular value for the three parameters $x_h, \gamma, n$. This provides, in particular,
the values of the fields and their derivatives at $x=x_h$ with the accuracy demanded.
Then a second integration is performed to determine the 
form of the solutions in the interior region
by using the data at $x=x_h$ as an initial value. 
 
As already stated, the symmetry (\ref{rescaling})
will be exploited  to set $x_h =1$.  This  choice of the scale
 allows for the two known limits $\gamma \to 0$ and $n \to 0$ to exist continuously.

\subsection{Case $n=0$}
The first problem  is to determine how the vacuum solution (\ref{vacuum})
is affected by the scalar field through the 
non-minimal coupling to the Gauss-Bonnet term. This was the object of \cite{Sotiriou:2014pfa}
but we briefly summarize this result for completeness.
Setting $\gamma \neq 0$, the  non-homogeneous part in the equation for the scalar field $\phi$ enforces this function to be non-trivial.
Remark that, since the initial Lagrangian depends only on the product $\gamma \phi$, only
 the case $\gamma > 0$ needs to be emphasized. 

\begin{figure}[ht!]
\begin{center}
{\label{c1}\includegraphics[width=8cm]{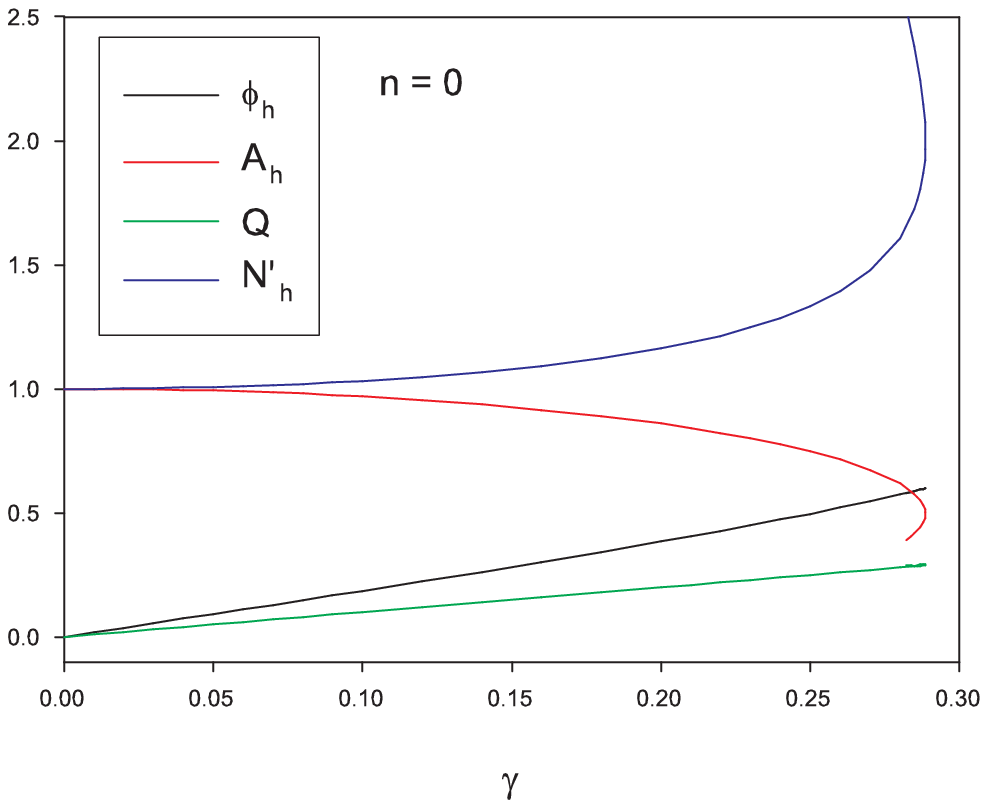}}
{\label{s0}\includegraphics[width=8cm]{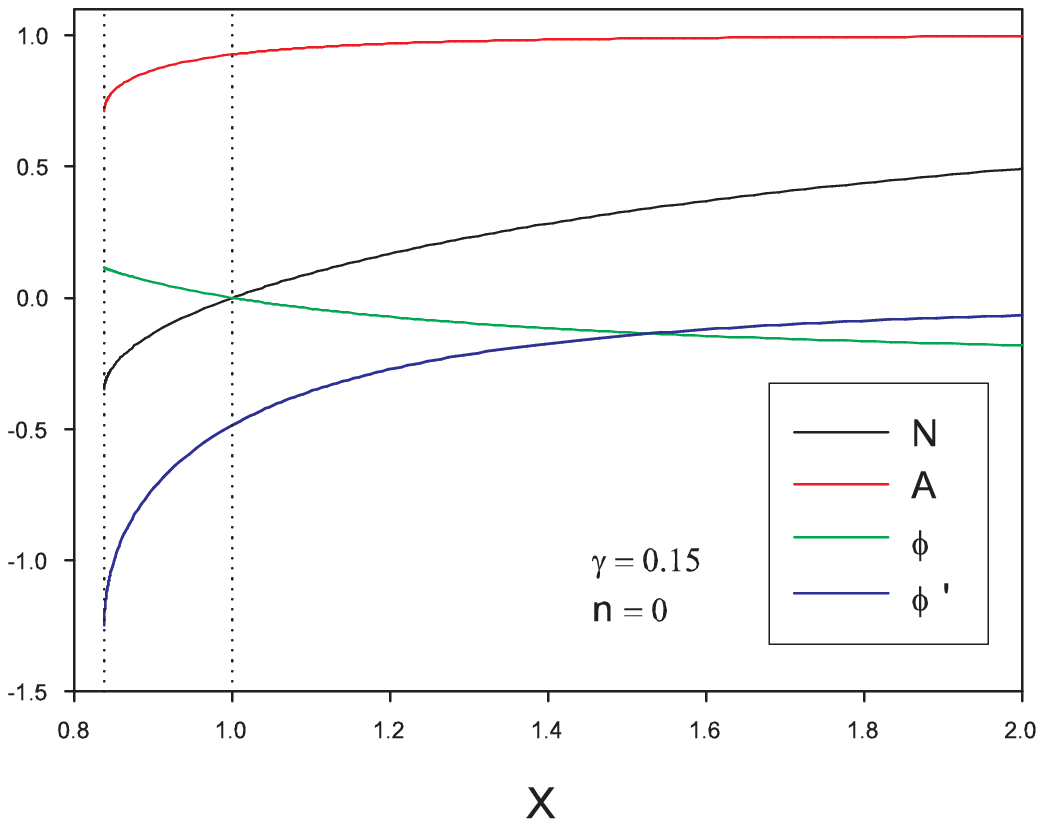}}
\caption{ {\it Left :} The dependence of some parameters  on the Gauss-Bonnet constant $\gamma$ for $n=0.0$ 
and $x_h=1.0$. 
{\it Right :} The profile of the solution with $\gamma = 0.15$ and $n=0$.
\label{gamma_vary}
}
\end{center}
\end{figure}

The corresponding ``Hairy-solution'' is characterized, namely, by the values 
$\phi'(x_h)$, $A(x_h)$, $N'(x_h)$, $\phi_{\infty}$ 
as well as  by  the mass $M$ and the charge $Q$. These parameters are determined numerically.
$M$ and $Q$ read off from the asymptotic decay of the fields  (\ref{asymptotic}).
Some of these parameters are reported as functions of $\gamma$ on the left side of  Fig.~\ref{gamma_vary}.  

In this case the regularity of the solutions on the horizon requires
\be 
            \phi'(x_h) = \frac{- x_h^2 \pm \sqrt{x_h^4-12 \gamma^2}}{2 x_h \gamma} \ \ , 
\ee
 implying in particular that real solutions will exist only for $\gamma \leq \gamma_{\max} =  1/\sqrt{12}$.  
The branch of solutions connecting to the vacuum in the limit $\gamma \to 0$ corresponds to the $+1$ sign.
A branch of solutions corresponding to the $-1$ sign exists as well and is represented
partly on the figure. It is very likely that this branch can be continued for smaller values of $\gamma$ but
the numerical computation of the second branch appeared to be  tricky, alterating the
numerical accuracy of the results. On the figure we limited the data of this branch to the values with a reliable accuracy.  

\begin{figure}[ht!]
\begin{center}
{\label{profile_1}\includegraphics[width=8cm]{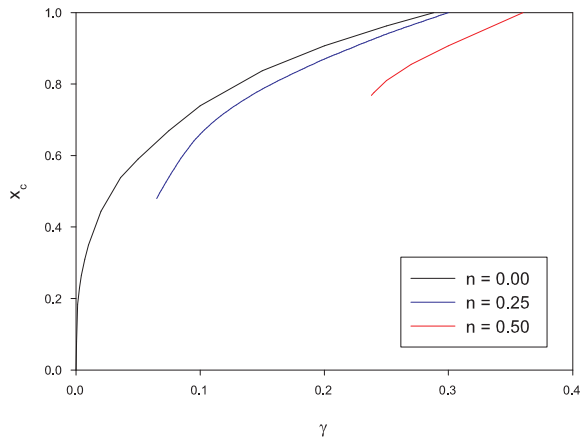}}
\end{center}
\caption{Dependence of the critical radius $x_c$ on $\gamma$ for several values of $n$.
\label{critical}
}
\end{figure}

The profile of the solutions for  $n=0$ and $\gamma= 0.15$ is presented on the right side of Fig.~\ref{gamma_vary}.
As pointed out in \cite{Sotiriou:2014pfa}, the integration of the equations in the interior region shows that
the metric and scalar functions are limited in a region $x_c < x < \infty$.
The  value $x_c$ corresponds to a  critical radius where the derivatives 
of the fields and -- by a consequence -- the metric invariants $R$ and $K$ diverges.
In the case of Fig.~\ref{gamma_vary}, we find  $x_c \approx 0.8375$. 
The dependence of $x_c$ on $\gamma$ was presented in \cite{Sotiriou:2014pfa}
but it is reproduced by means of the black line  on Fig.~\ref{critical} for the sake of comparison with the case $n > 0$.
\subsection{Case $n >0$}
We now discuss  the solutions obtained  for $n > 0$.

\textbf{Exterior :} In the exterior region, the family of solutions
corresponding to a fixed $n$ and varying $\gamma$ present qualitatively the same features as in the case
$n=0$. One quantitative  difference with respect to $n=0$ solutions  is related to the fact that
the condition of regularity (\ref{regularity}) now depends on the value $A(x_h)$. Since 
this is determined numerically,   the value $\gamma_{\max}$
cannot be determined analytically for $n > 0$. The value
 $\gamma_{\max}$ increases slightly with $n$; we have $\gamma_{\max}=1/\sqrt{12} \approx 0.2886$ for $n=0$ and we find
 $\gamma_{\max} \approx 0.3$ and  $\gamma_{\max} \approx 0.36$ respectively for $n=0.25$ and $n=0.50$,
  as sketched on Fig.~\ref{critical}. The value $\gamma_{\max}$ corresponding to the value of $\gamma$ for which the critical radius $x_c$ becomes equal to the horizon radius $x_h$ (see discussion below). The solution then stop existing before exhibiting a naked singularity.
  
  Note~: the reason the lines do not reach $\gamma = 0$ for $n>0$ will be discussed below.

\textbf{Interior :} Because the vacuum (\ref{vacuum}) NUT solution is everywhere regular in the interior region $r < r_h$,
the question of the structure of the  Nutty-hairy black holes for $r < r_h$ raises naturally.  
The pattern is  not so simple due to a  double critical phenomenon which we now explain.

\begin{figure}[ht!]
\begin{center}
{\label{c1}\includegraphics[width=8cm]{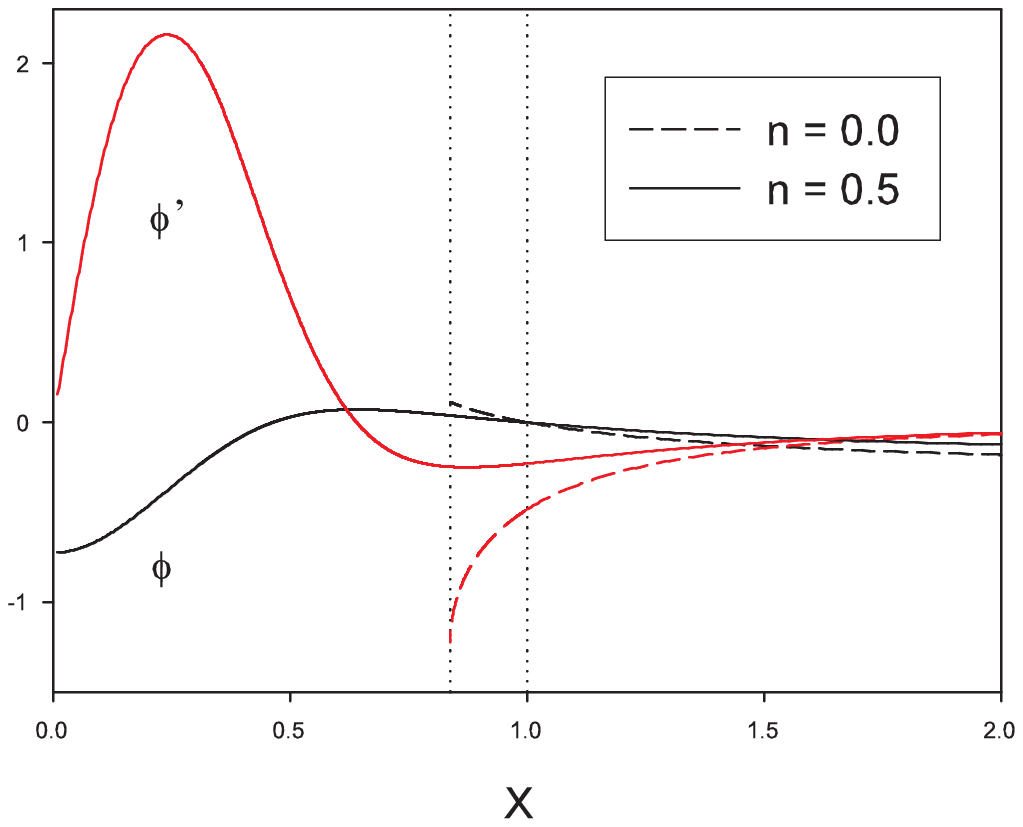}}
{\label{s0}\includegraphics[width=8cm]{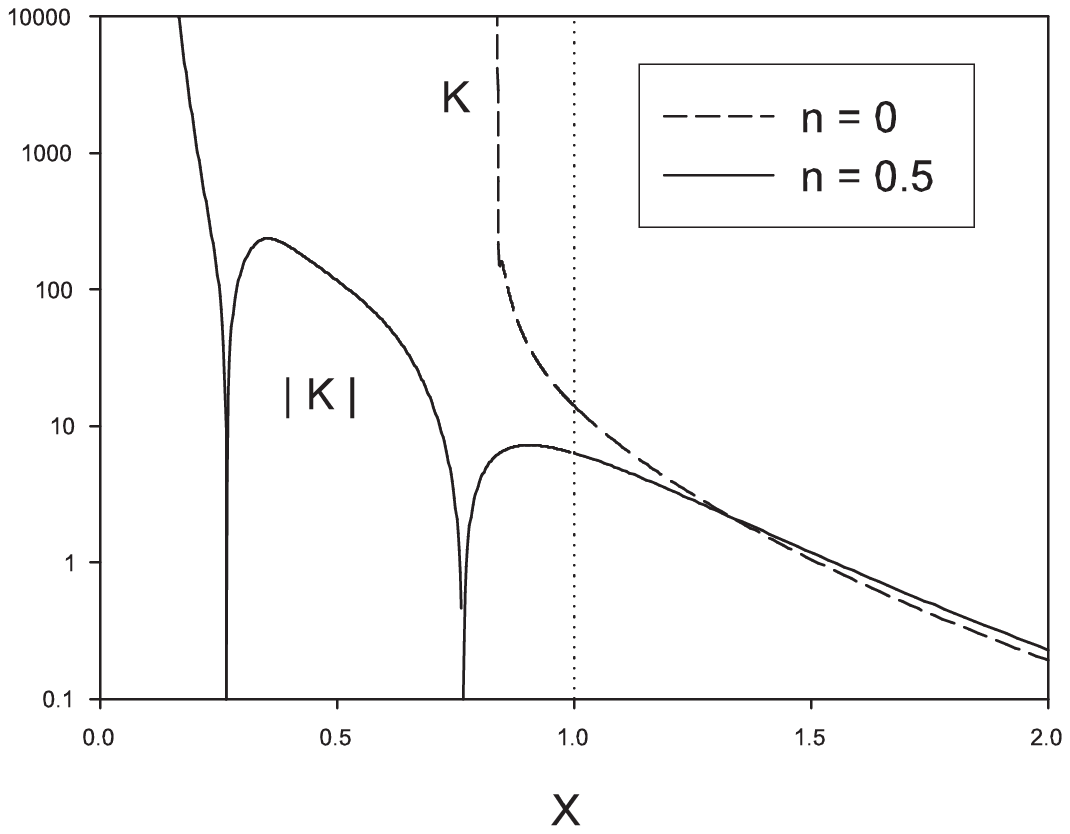}}
\caption{ {\it Left}: The profiles of the scalar field $\phi$ and its derivative $\phi'$ for $\gamma=0.15$
and two values of $n$. 
{\it Right:}
The corresponding Kretschmann  scalar $K$. Note the spikes of the solid line correspond to zeros of $K$ because of the logarithmic scale on the $y$ axis.
\label{profile_n_05}
}
\end{center}
\end{figure}

Fixing $n>0$ we know that the vacuum solution (\ref{vacuum})
is regular for $0 \le x < \infty$. Increasing the coupling constant gradually $\gamma$, with  $n>0$ fixed,
the numerical results strongly suggest that
the  hairy-nutty-black hole is  regular for $0 < x < \infty$. 
In the limit $x \to 0$, our results indicate in particular  $N(x) \to - \infty$
while the scalar field remains finite in agreement with the perturbative expression (\ref{perturbation}).
The corresponding Ricci scalar (and the other invariants) also diverge to infinity, 
confirming the occurrence of a singularity at the origin. This is illustrated on Fig.~\ref{profile_n_05} where the
profiles of the solutions corresponding to $\gamma=0.15$ are superposed for $n=0$ (dashed lines) and $n=0.5$ (solid lines).

We now discuss the second singularity: again with $n$ fixed, an increase of $\gamma$ reveals another
peculiarity of the nutty black holes  in the interior region. The results indicate that,
apart from the singularity at the origin,
a second singularity appears at an intermediate radius, say at $x = x_c$ (with    $0 < x_c < x_h$)
when the Gauss-Bonnet coupling constant approaches a critical value, $\gamma \to  \gamma_c$.
For example, for $n=0.25$ and $n=0.5$, we find respectively $\gamma_c \approx 0.065$ and $\gamma_c \approx 0.238$~;
this explains why the lines on Fig.~\ref{critical} stop suddenly without reaching $\gamma = 0$.

\begin{figure}[ht!]
\begin{center}
{\label{profile_2}\includegraphics[width=8cm]{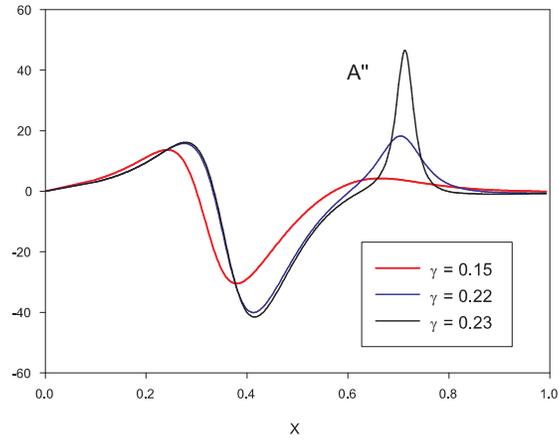}}
\end{center}
\caption{Profile of the second derivative $A''$ for $n=0.5$ and three values of $\gamma$.
\label{a_bis_interior}
}
\end{figure}

This phenomenon is not so  easy to detect from the numerical solutions and  
 is manifest only through a careful examination of the derivatives
of the metric fields $A$ and $N$. As an  illustration we plot on   Fig.~\ref{a_bis_interior}.
the function $A''$ for  $n= 0.5$ and for three values of $\gamma$  approaching $\gamma_c$.
In this case we find 
$\gamma_c \approx 0.238$, with the corresponding critical radius $x_c\left(\gamma_c\right) \approx 0.768$, and $\gamma_{\max} \approx 0.36$.
One can appreciate on this plot that, while $\gamma$ approach $\gamma_{c}$, a ``bump'' appears in the profile of $A''$. This bump grows quickly -- and possibly tend to infinity, although it is hard to verify numerically -- in the limit $\gamma \to \gamma_{c}$ and is roughly centred around $x_c(\gamma_{c})$.
For $\gamma > \gamma_c$, the pattern of \cite{Sotiriou:2014pfa} is recovered~: the solutions can be continued only
for $0 < x_c < \infty$. As illustrated on Fig.~\ref{critical}, $x_c$ increases with $\gamma$ and the solution stops to exist in the limit $\gamma \to \gamma_{\max}$ for which $x_c\left(\gamma \to \gamma_{\max}\right) \to x_h$. Then, as we discussed above, the solution stops before reaching a naked singularity.
\clearpage

\section{Geodesics around hairy-nutty black holes}\label{light-geodesics}

In this section we study the  geodesics in a nutty-black hole space-time with a special emphasis
on the effects of the NUT charge and of the Gauss-Bonnet coupling term. 
We first describe the generic properties of the curves and then study the possible motions in the equatorial plane.

\subsection{Equations and constants}

We parametrize the geodesics by means of the functions
$$X(\lambda) \ = \left(T(\lambda),R(\lambda),\Theta(\lambda), \Phi(\lambda)\right)~,$$ 
where $\lambda$ is an affine parameter.
The equations of geodesics in the space-time of a black hole such as constructed in the previous section 
can be set as follows~: 
{\small{\begin{align*}
\label{full_system}
  \ddot{T}(\lambda) &= \frac{-2\dot{A}\dot{R}(\dot{T}-2n(\cos(\Theta)-1)\dot{\Phi})}{A}+\frac{\dot{N}\dot{R}(2n(\cos(\Theta)-1)\dot{\Phi} -\dot{T})}{N} 
  \\ & +\frac{n}{\Sigma}\left( -4\sin^2(\frac{\Theta}{2}) \left\{-2R\dot{R}+\Sigma \tan(\frac{\Theta}{2})\dot{\Theta} \right\}\dot{\Phi} \right. \\
  & \left.  - 4nA^2N\dot{\Theta} \left\{\tan(\frac{\Theta}{2}) \dot{T} +8n\csc(\Theta)\sin^4(\frac{\Theta}{2})\dot{\Phi}\right\} \right)~, \\
  \ddot{R}(\lambda) &= -A\dot{A}N^2\Delta + NR(\dot{\Theta}^2 + \sin^2(\Theta)\dot{\Phi}^2) - \frac{\dot{N}(A^2N^2\Delta^2-\dot{R}^2)}{2N}~, \\
  \Sigma  \ddot{\Theta}(\lambda) &= -2R\dot{R}\dot{\Theta} - 2nA^2N\sin(\Theta)\dot{T}\dot{\Phi} + \dot{\Phi}^2 \left[\Sigma \sin(\Theta)\cos(\Theta)+n^2A^2N(2\sin(2\Theta)-4\sin(\Theta))\right]~, \\
  \Sigma \ddot{\Phi}(\lambda) &= -2R\dot{R}\dot{\Phi} + \dot{\Theta}\left[ 2\cot(\Theta)\Sigma\dot{\Phi} + nA^2N\left\{2\csc(\Theta)\dot{T}-(4n(\cot\Theta-\csc\Theta)\dot{\Phi}) \right\}\right]~,
\end{align*}}}
where $\dot{f} = \frac{\diff f}{\diff \lambda}~$
and $$\Delta \equiv  \left(\dot{T}-2n\dot{\Phi}\left(\cos\Theta-1\right)\right) \ \ , \ \ \Sigma \equiv n^2 + R^2 \ .$$
The metric functions $N(R(\lambda)), A(R(\lambda)), \cdots$ are obtained numerically (see the previous sections).

The above  equations are solved with the initial conditions: 
\begin{equation}
\label{initial}
 X(\lambda = 0) = (0, r_0, \theta_0,\varphi_0) \, , \, \dot{X}(\lambda = 0) = (k_t, k_r, k_\theta, k_\varphi), 
\end{equation}
where $k_t$ is related to the energy of the particle along the geodesic, $k_r$ is essentially its radial velocity, and $k_\theta, k_\varphi$ are angular velocities. 

The quadrivector  $\dot{X}$ is subject to the the condition $\dot{X}^\mu g_{\mu\nu} \dot{X}^\nu = - \epsilon$
with $\epsilon =0,1,-1$ respectively for lightlike, time-like and space-like geodesics. Note that $\dot{X}$ is nothing else than the wave vector when the affine parameter is the proper time along the geodesic. The constraint fixes the energy of the particle once the 3-momentum is fixed.

The case $\epsilon = 0$ describes the propagation of light rays and the case $\epsilon = 1$ corresponds to massive particles. The case $\epsilon = -1$, corresponding to tachyonic motions, is unphysical and would not be considered in the following. 

Using the fact that $\partial_t$ and $\partial_\varphi$ are Killing vectors, one can find two constants of motion along the geodesics. Denoting $\dot{X} = \left(\dot{T},\dot{R},\dot{\Theta}, \dot{\Phi}\right)$ the 4-velocity along a geodesic, the constants obtained with $\partial_t$ and $\partial_\varphi$ respectively read :
\begin{align}
 E & = \left(\partial_t\right)_\mu \dot{X}^\mu = g_{0 \mu} \dot{X}^\mu = - A^2(R) N(R) \left(\dot{T} + 4 n \sin^2\left(\frac{\Theta }{2}\right) \dot{\Phi}\right)~, \\
 L & = - \left(\partial_\varphi\right)_\mu \dot{X}^\mu = - g_{3 \mu} \dot{X}^\mu \\ &
 = 4 n \sin ^2\left(\frac{\Theta}{2}\right) A^2(R) N(R) \dot{T} + \left(16 n^2 A^2(R) \sin ^4\left(\frac{\Theta}{2}\right) N(R) - \sin ^2(\Theta) \left(n^2+R^2\right)\right) \dot{\Phi}. \notag
\end{align}
The constant $E$ might be interpreted as the energy of the particle moving along the geodesic and $L$ as an analogue to the third component of the angular momentum of the particle.
The above relations can then be used to express
the quantities $\dot{T}$ and $\dot{\Phi}$ in terms of the functions  $R, \Theta$  
and the constants  $E,L$ :
\begin{align}\label{Tprim}
 \dot{T} &  = \frac{n \left(\sec ^2\left(\frac{\Theta }{2}\right) (L + 4 n E)-4 n E\right)}{n^2+R^2}-\frac{E}{A^2(R) N(R)}~, \\
 \dot{\Phi} & = -\frac{\csc ^2(\Theta ) (L + 2 n E -2 n E \cos (\Theta ))}{n^2+R^2} \ \ . \label{Phiprim}
\end{align}
As  a consequence, the system of four equations above reduces to the two equations
corresponding to the functions
 $R$ and $\Theta$.
These  equations are  lengthy and we do not write them here.
\subsection{Generic motions}
\label{subsectgenmotion}
\begin{figure}[h!]
 \begin{minipage}[c]{.48\linewidth}
  \begin{overpic}[width=0.9\textwidth]{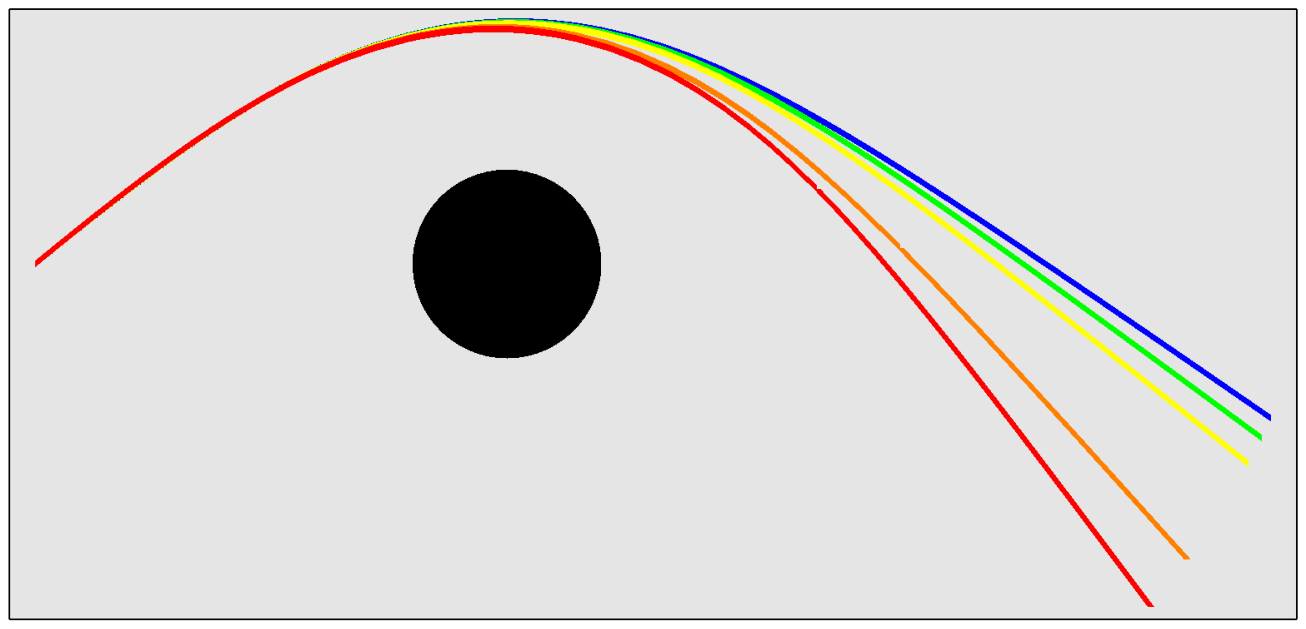}
   \put (6,22) {$\displaystyle\mathcal{O}$}
  \end{overpic}
 \end{minipage} \hfill
 \begin{minipage}[c]{.48\linewidth}
  \begin{overpic}[width=0.9\textwidth]{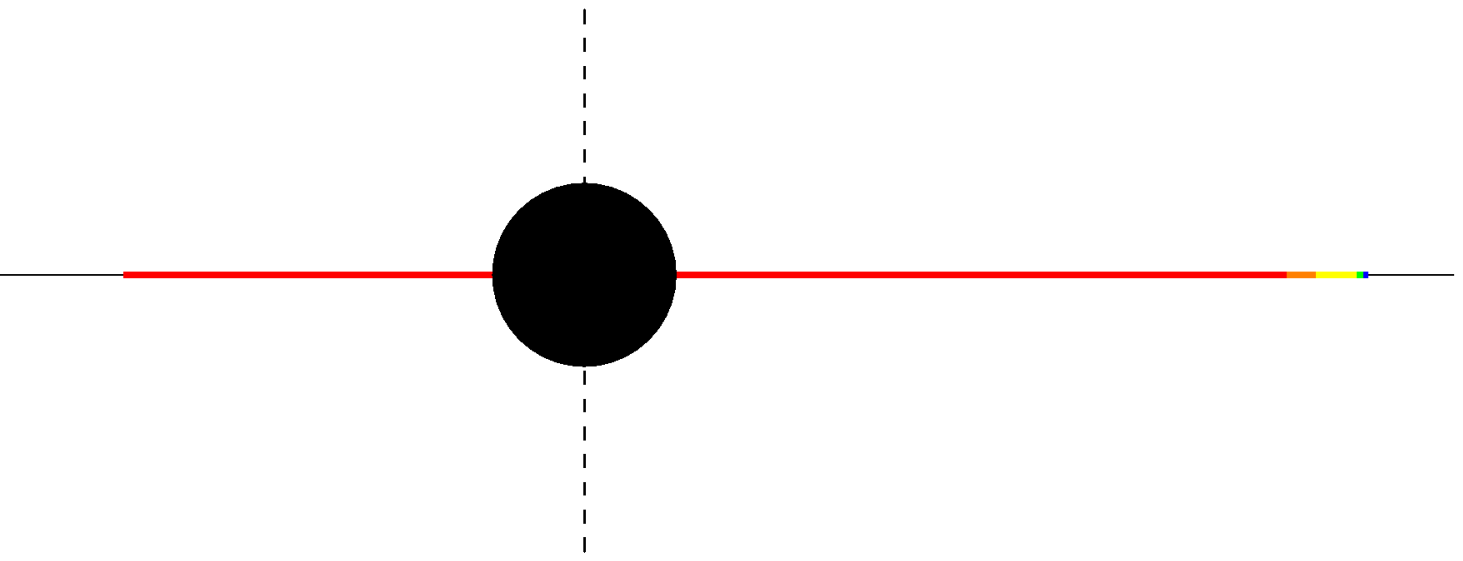}
   \put (6,17) {$\displaystyle\mathcal{O}$}
  \end{overpic}
 \end{minipage}
 
 \begin{minipage}[c]{.48\linewidth}
  \begin{overpic}[width=0.9\textwidth]{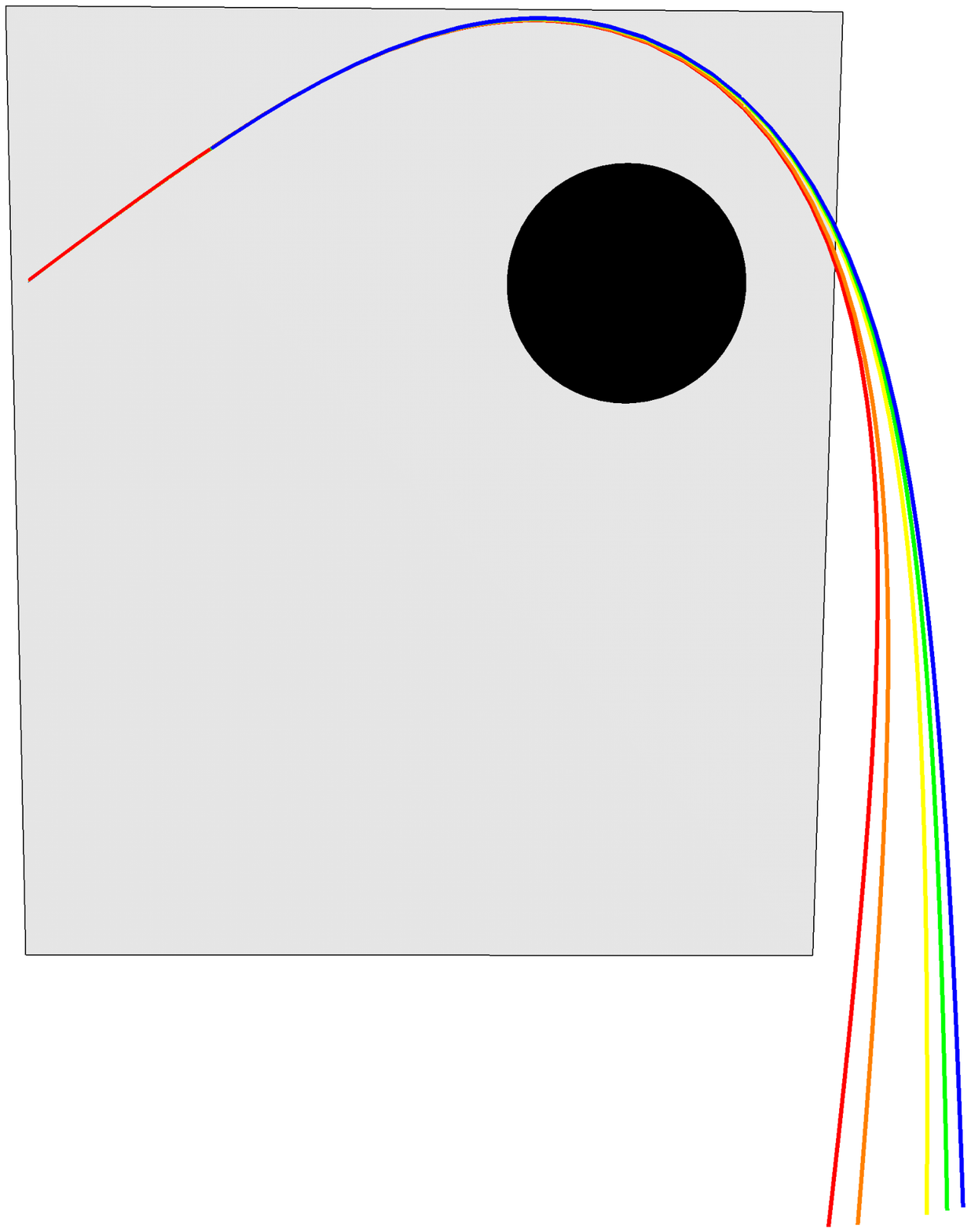}
   \put (15,68) {$\displaystyle\mathcal{O}$}
  \end{overpic}
 \end{minipage} \hfill
 \begin{minipage}[c]{.48\linewidth}
  \begin{overpic}[width=0.9\textwidth]{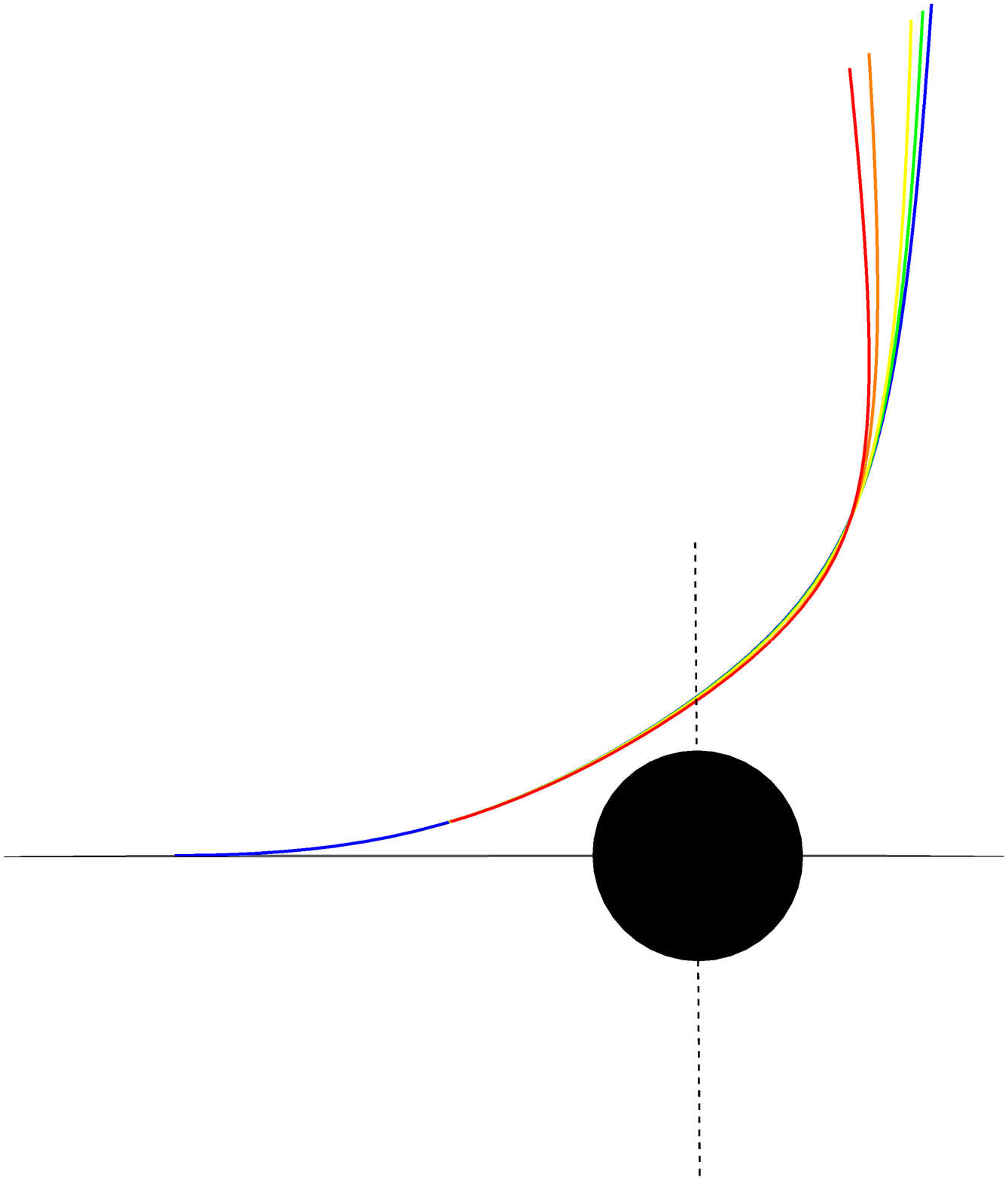}
   \put (20,29) {$\displaystyle\mathcal{O}$}
  \end{overpic}
 \end{minipage}
 
 \caption{\small{\textit{Top (Left and Right)} : $n=0$. \textit{Bottom (Left and Right)} : $n=1$. In the two cases, the considered values of the coupling constant are $\gamma = 0$, $\gamma = 0.1$, $\gamma = 0.15$, $\gamma = 0.25$, $\gamma = 0.28$ , (respectively in blue, green, yellow, orange and red). The black sphere corresponds to the horizon $r=r_h=1$. The observer is located at a distance $r=5 r_h$ and lies in the equatorial plane. It's position is denoted by $\mathcal{O}$. The photons are sent with identical starting values, so their different geodesics are only function of $\gamma$.}}
 \label{vary_gamma}
\end{figure}

We haved solved numerically the geodesic equations for several values
of the parameters $n, \gamma$ and of the initial conditions (\ref{initial}).
Let us  first present  families  of geodesics highlighting the effects of the Gauss-Bonnet parameter $\gamma$ and 
of the NUT charge $n$. 
Examples of light-like geodesics with fixed wave-vector  are shown in Fig.~\ref{vary_gamma}
for two values of $n$ and several values of $\gamma$.

Let us first discuss the {\bf top} part of Fig.~\ref{vary_gamma} corresponding to  $n=0$.
On this figure, we show the same set of geodesics seen from  different points of view :
the left side represents  the equatorial  (or $XY$) plane, in grey, while the right side  represents the YZ plane 
with the $OZ$ axis figured out by the dashed line.   
We see in particular that the photons evolve in a plane for all values of $\gamma$. The Gauss-Bonnet coupling constant just changes the curvature of the different lines; increasing $\gamma$ the black hole becomes ``more and more attractive'' since the lines become more and more curved. 

The {\bf bottom} part of Fig.~\ref{vary_gamma} corresponds to  $n=1.0$.
Here the trajectories cease to be planar. This could be expected since, turning on the NUT parameter one also turns on the $(t\varphi)$ component of the metric. Consequently, for $n \neq 0$, we deal with a stationary non-static space-time. This case is similar to the case of a rotating black hole for which frame-dragging effects are well known. 

\begin{figure}[H]
 \centering
 \includegraphics[scale=0.35]{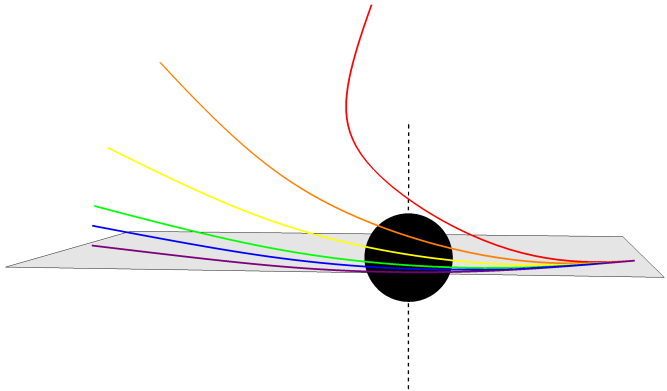} 
 \caption{Geodesics for $\gamma = 0$ and for $n=0$, $n=0.05$, $n=0.1$, $n=0.25$, $n=0.5$, $n=1$, respectively represented in purple, blue, green, yellow, orange and red. The setup is identical as in figure \ref{vary_gamma}.}
 \label{vary_n}
\end{figure}

Fig.~\ref{vary_n}  confirms  that trajectories do not lie in a plane for $n\neq 0$ and 
that increasing the NUT parameter causes an increase of the geodesics curvature and torsion 
(defined as usual for curves).

\begin{figure}[H]
 \begin{minipage}[c]{.48\linewidth}
  \includegraphics[scale=0.6,trim={1cm 10cm 1cm 10cm},clip]{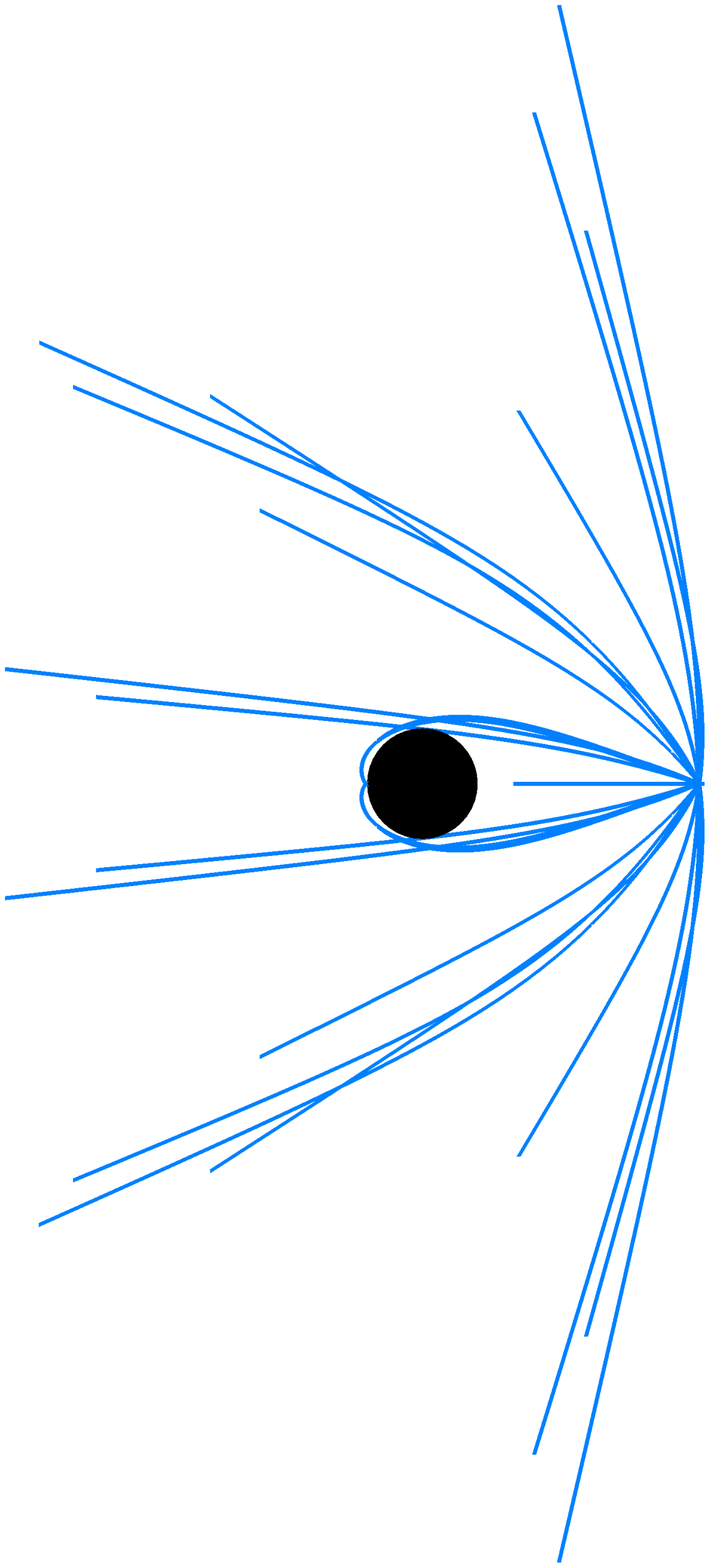}
 \end{minipage} \hfill
 \begin{minipage}[c]{.48\linewidth}
  \includegraphics[width=\textwidth,trim={3cm 3cm 3cm 3cm},clip]{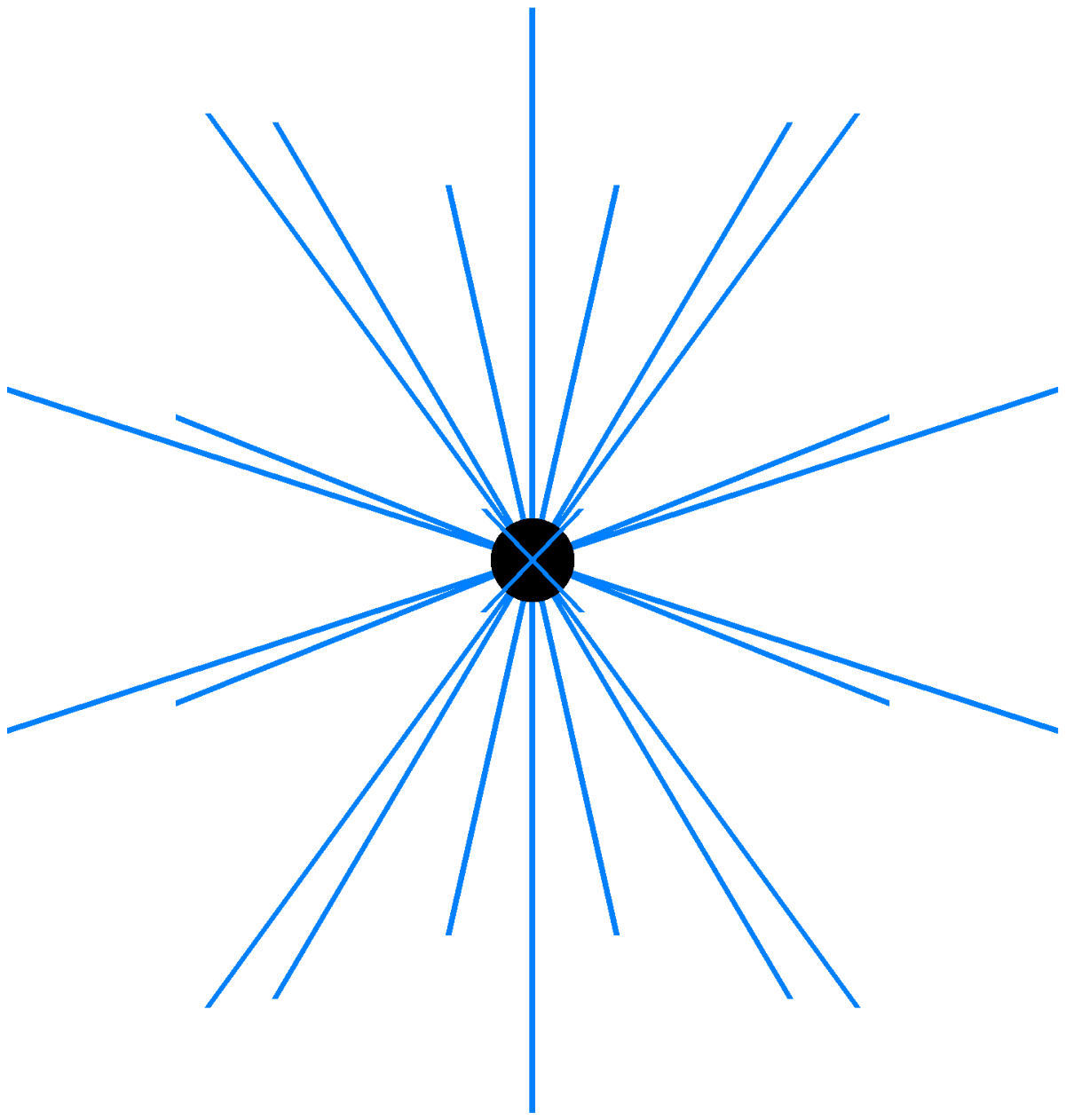}
 \end{minipage}
 \begin{minipage}[c]{.48\linewidth}
  \includegraphics[scale=0.6,trim={1cm 10cm 1cm 10cm},clip]{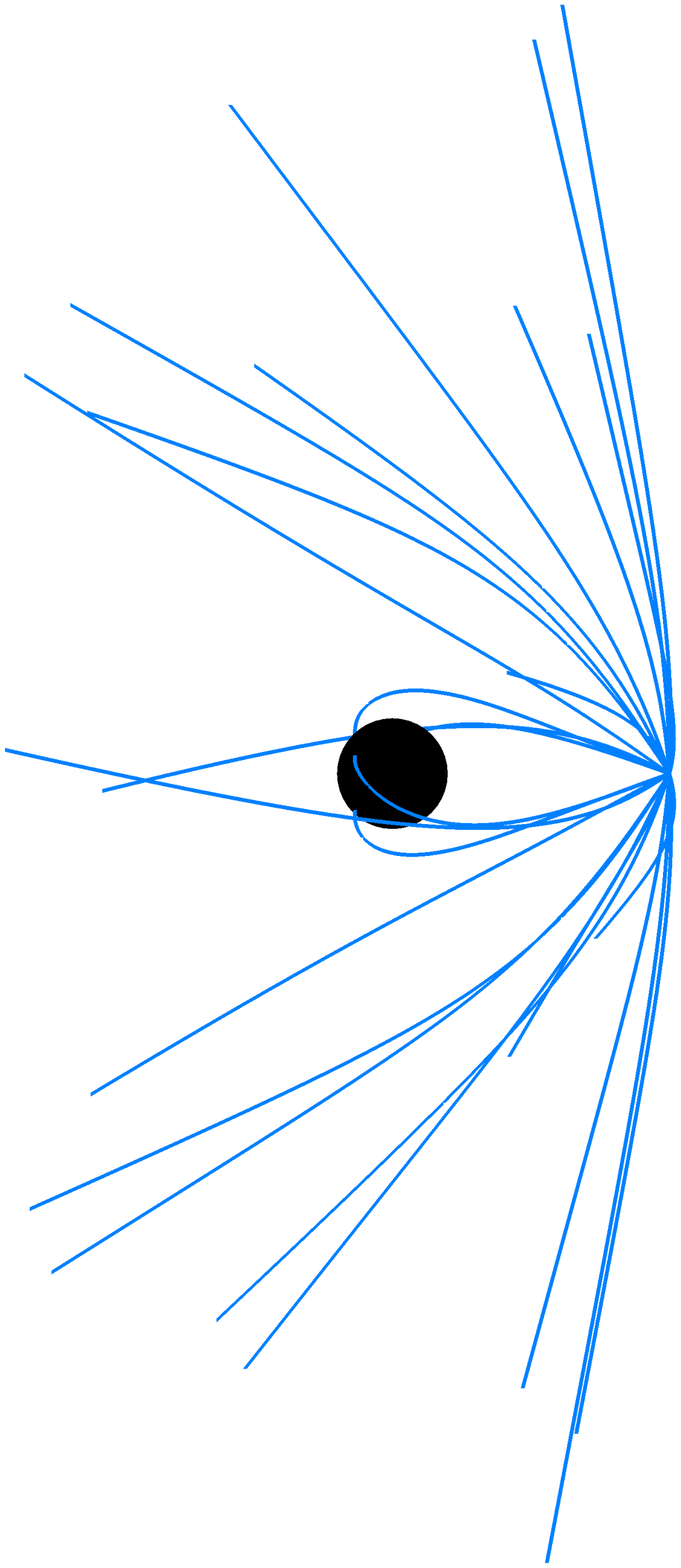}
 \end{minipage} \hfill
 \begin{minipage}[c]{.48\linewidth}
  \includegraphics[width=\textwidth,trim={3cm 3cm 3cm 3cm},clip]{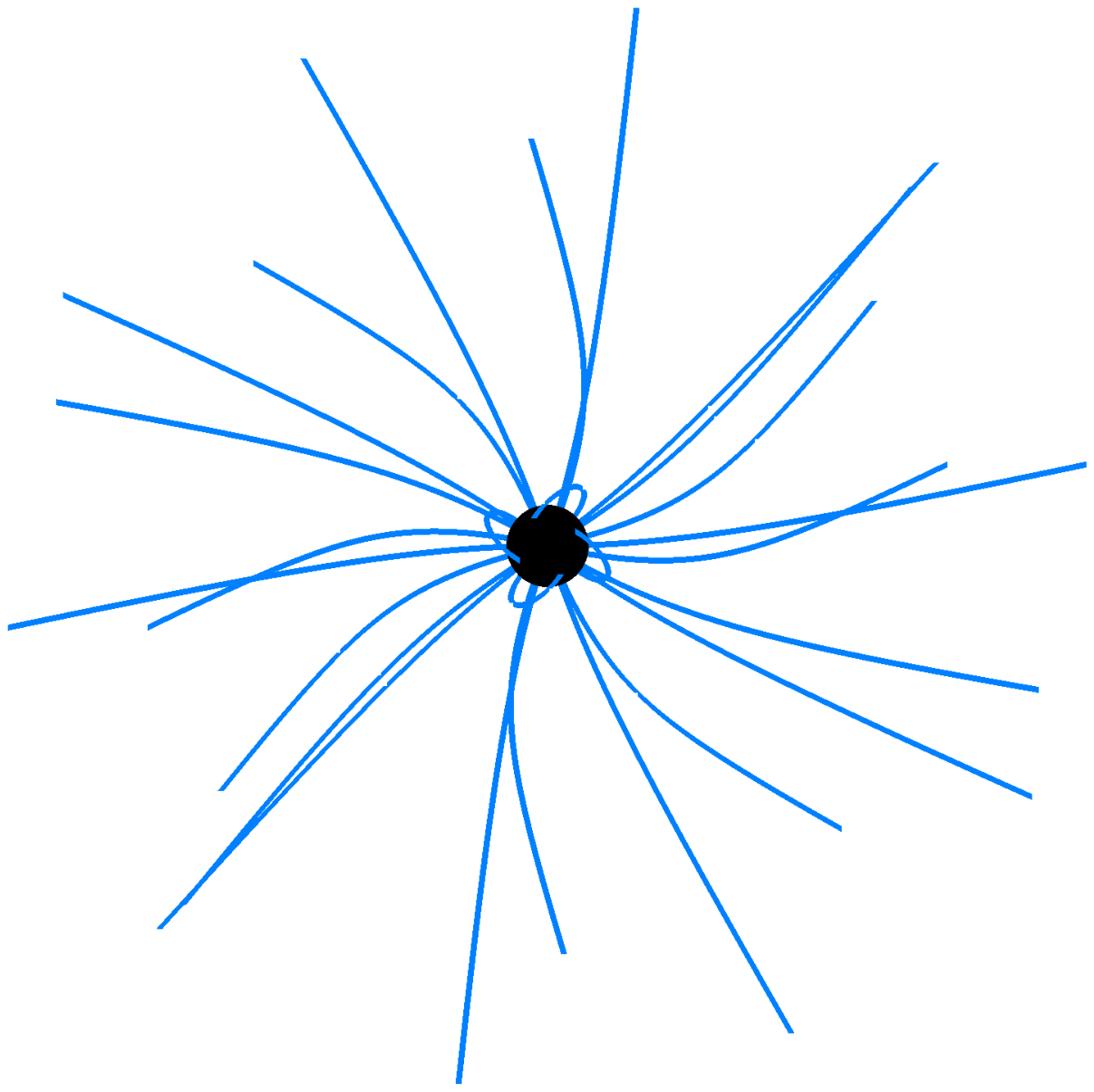}
 \end{minipage}
 \caption{\small{\textit{Top (left and right)} : $n=0$. \textit{Bottom (left and right)} : $n=1$. Light rays for various initial directions and $\gamma = 0.1$ are represented. The observer lies where the trajectories meet. On the top and bottom figures, the left and right situations are the same, but from a different point of view so the behaviour of the geodesics becomes clearer.}}
 \label{spiders}
\end{figure}

The above statement is further illustrated in Fig.~\ref{spiders} where  various trajectories are shown for two values of $n$ and  $\gamma = 0.1$. For $n \neq 0$, the frame-dragging effect  is clearly seen on the right part of the figure.
Let us highlight that the purpose of this figure is to reveal the global frame-dragging feature
rather than quantitative details. 
In order to be as exhaustive as possible, analogous results for other setups are shown in the appendix.

\subsection{Motions in the equatorial plane}

We now investigate the possibility of geodesic motion in the equatorial plane 
(i.e. with $\Theta(\lambda) = \frac{\pi}{2}, \forall \lambda$). 
For this purpose, we  fix as  initial conditions $\Theta(0) = \frac{\pi}{2}$ and $\dot{\Theta}(0) = 0$.
The relevant conditions to guarantee that these initial conditions will lead to a constant value of $\Theta$ for all values of $\lambda$ are then obtained through the equation fixing the  $\ddot{\Theta}$ function
which  turns out to be~:
$$\ddot{\Theta}(\lambda) = - \frac{2 E n (L + 2 n E)}{\left(n^2+R(\lambda)^2\right)^2} \ .$$

Then, a given geodesic would stay in the equatorial plane \textbf{iff} $\ddot{\Theta}(\lambda) = 0$ for all $\lambda$, namely \textbf{iff} 
\begin{equation}
 \label{condequator}
 n = 0 \lor E = 0 \lor L_3 = 0,
\end{equation}
where $L_3 \equiv L + 2 n E$.

The planarity of trajectories for $n=0$ was already pointed out~; the other two solutions, 
which somehow are \textit{a priori} unexpected, are worth being examined. 
In order to understand what happens in these cases, we have to examine the last equation of geodesics : the equation of the $R$ function. 
One can look directly at the $\ddot{R}$ equation or, equivalently, use the condition $\dot{X}^\mu g_{\mu\nu} \dot{X}^\nu = - \epsilon$  together with $\Theta(\lambda) = \frac{\pi}{2}$ and the equations \eqref{Tprim} and \eqref{Phiprim} to obtain : 
\begin{equation}
 \label{rpointcarre}
 \dot{R}(\lambda)^2 = U\left(R; n, E, L_3\right)~,
\end{equation}
where
\begin{equation}
 U\left(R; n, E, L_3\right) = \frac{E^2}{A(R)^2} - N(R) \left(\frac{(L_3)^2}{n^2+R^2} + \epsilon\right).
\end{equation}

The  equation above can be studied as a potential-like equation : the motion is only possible in the regions where $U\left(R; n, E, L_3\right) \ge 0$ so a study of the properties of the ``potential'' $U\left(R; n, E, L_3\right)$ will give us all the necessary information to classify the geodesics living in the equatorial plane. Let us emphasize that Eq.~\eqref{rpointcarre} is relevant \textbf{iff} $(n = 0 \lor E = 0 \lor L_3 = 0)$. 

\begin{figure}[H]
 \begin{minipage}[c]{.48\linewidth}
  \includegraphics[scale=0.6]{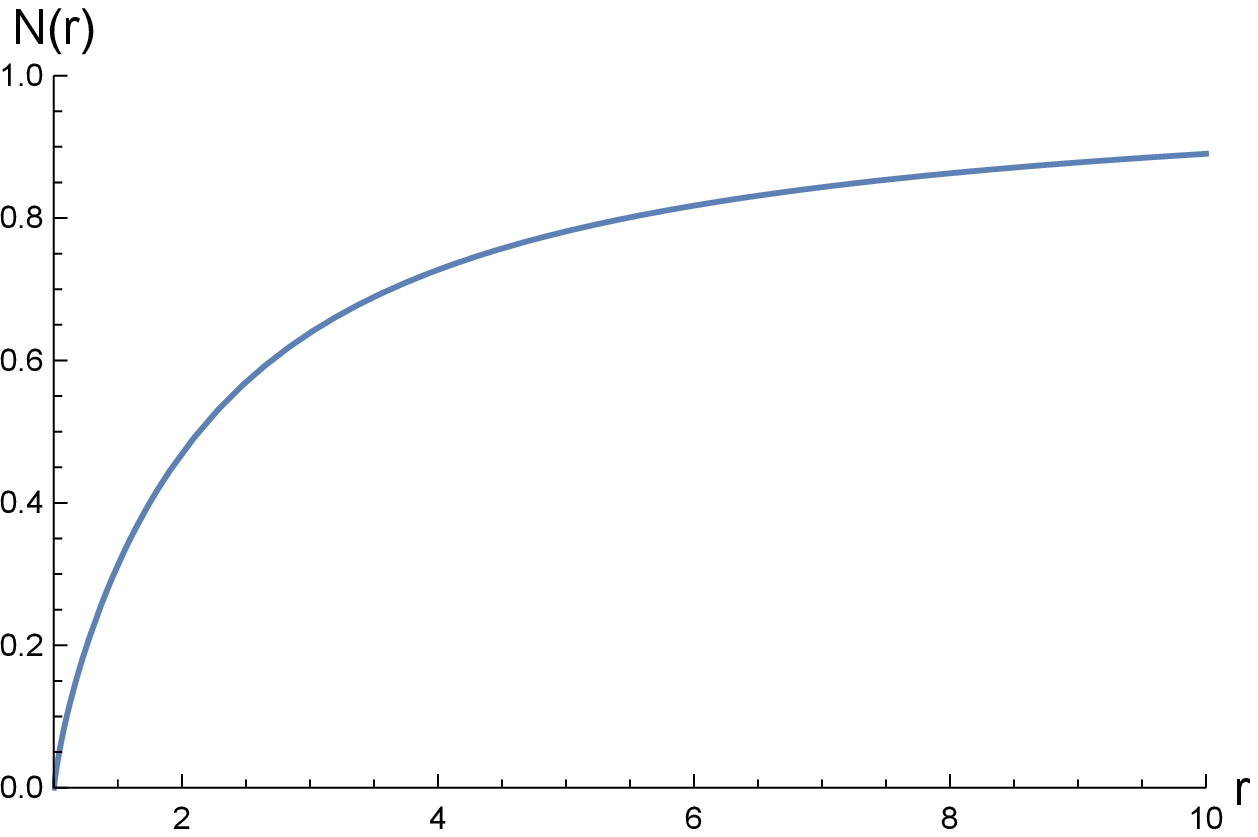}
 \end{minipage} \hfill
 \begin{minipage}[c]{.48\linewidth}
  \includegraphics[scale=0.6]{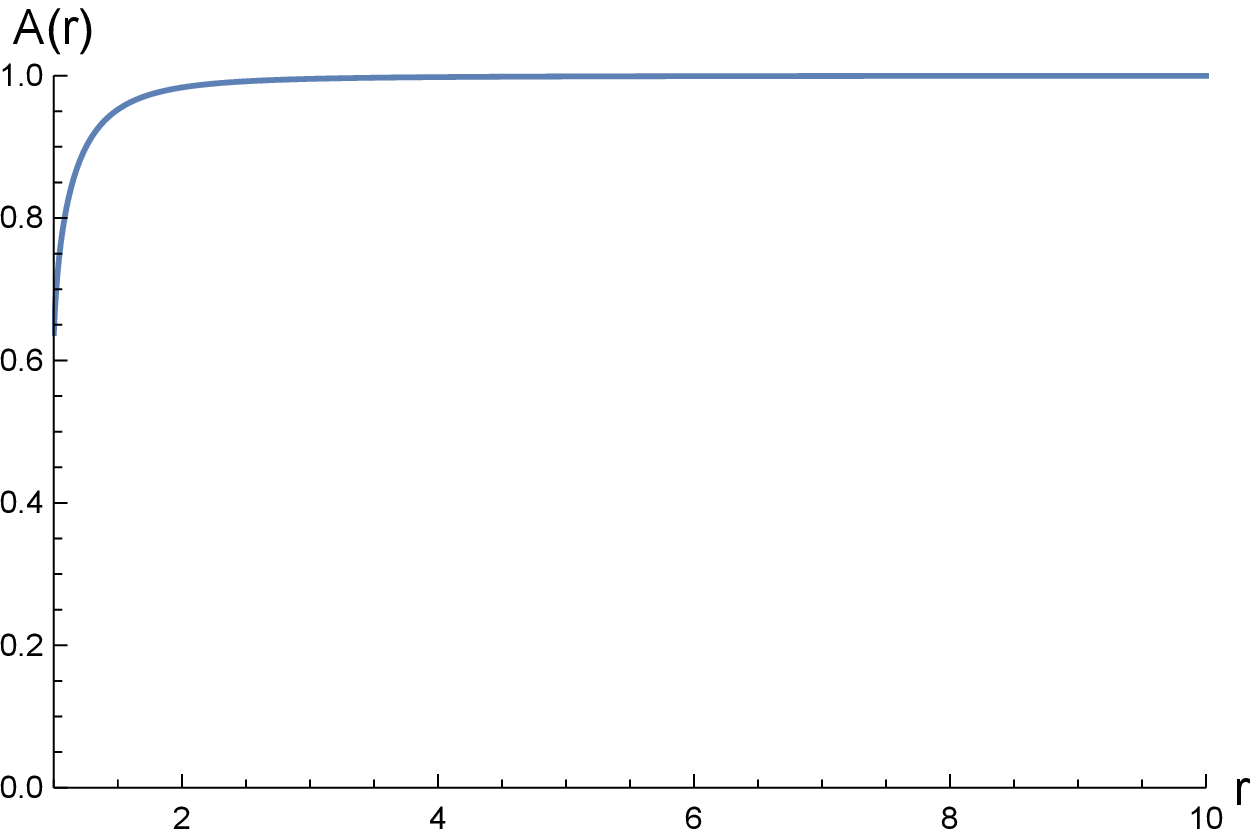}
 \end{minipage}
 \caption{Profile of the metric functions $N$ (left side) and $A$ (right side) in the exterior region for $n = 0.1$ and $\gamma = 0.28$. Here $A(r_h) = 0.639$ and $A(2 r_h) = 0.983$ ($r_h = 1$). For the sake of comparison, we used the same scale on both plots.}
 \label{figNA}
\end{figure}

As a consistency check, note that when $n = 0$ and $\gamma = 0$, the situation should reduce to the Schwarzschild scenario~; that is $A(r) = 1$ and $N(r) = 1 - \frac{2 M}{r}$. Using those expressions for $A$ and $N$, one can easily verify that \eqref{rpointcarre} reduces to the equation for geodesics in Schwarzschild background, see for example \cite{Riazuelo:2015shp}.
The influence of $n$ and $\gamma$ on $U(R; n, E, L_3)$ is mostly ``\emph{hidden}'' in the metric functions $A$ and $N$ -- especially for $\gamma$ since one should obtain $A$ and $N$ numerically for $\gamma \neq 0$. Actually, according to our numerical results, the behaviour of these functions remains more or less the same for all values of $n$ and $\gamma$ : 

As pointed out in Sect.~\ref{sectbound}, we have constructed our solutions such that $N(r_h) = 0$, $N(r) \underset{r\to\infty}{\longrightarrow} 1$ and $A(r) \underset{r\to\infty}{\longrightarrow} 1$. For all values of $n$ and $\gamma$, it turns out that $1 \ge A(r_h) > 0$ and that $N$ is a strictly increasing function on the exterior space-time (i.e. for $r\ge r_h$). $N$ would then smoothly grow from $0$ to $1$ as $r$ increase from $r_h$ to infinity (see Fig.~\ref{figNA} left side).

The situation for the function $A$ is a bit different. For $\gamma = 0$, $A$ is constant (and then $A(r)=1$ according to the boundary conditions). For $\gamma \neq 0$, $A$ acquires a non-trivial behaviour and becomes, as $N$, a strictly increasing function, growing from $A(r_h)$ to $1$ as $r$ increase from $r_h$ to infinity. However, unlike $N$, $A$ tends extremely quickly to $1$ (see Fig.~\ref{figNA} right side). Our numeric indicates that, for all values of $n$ and $\gamma$, one typically have $A(2 r_h) > 0.9$. Then, as a good approximation, $A(r) \approx 1$ for all values of $n$ and $\gamma$.

As a consequence, we have
$$ \infty > U\left(r_h; n, E, L_3\right) = \frac{E^2}{A(r_h)^2} > 0~,$$
and 
$$U\left(R; n, E, L_3\right) \approx E^2 - N(R) \left(\frac{(L_3)^2}{n^2+R^2} + \epsilon\right)\equiv E^2 - V(R).$$ 

Then the shape of the curve is due to $V(R)$ and the energy acts like a shift constant.

Since the relevance of \eqref{rpointcarre} require condition \eqref{condequator} to be satisfied, three cases might appear :

\subsubsection{Case $E = 0$ :}

In this case the ``potential'' reduces to
$$U\left(R; n, 0, L_3\right) = - N(R) \left(\frac{(L_3)^2}{n^2+R^2} + \epsilon\right) \le 0,$$
so there is no possible motion. Note that this case is twice disfavoured since, imposing $E = 0$, motions are only possible with $L_3 = L \neq 0$. Indeed, when $E = 0 = L$, Eq.\eqref{Tprim} would lead to $\dot{T} = 0$~; this would correspond to a particle which does not propagate in time.

\subsubsection{Case $L_3 = 0$ :}

This case corresponds to purely radial motion since, together with $\Theta(\lambda) = \frac{\pi}{2}$, $L_3 = 0$ lead to $\dot{\Phi} = 0$ via \eqref{Phiprim}. The ``potential'' is
$$U\left(R; n, E, 0\right) = \frac{E^2}{A(R)^2} - N(R) \epsilon~,$$
and its derivative with respect to $R$
$$U'\left(R; n, E, 0\right) \approx - N'(R) \epsilon.$$
\vspace{4mm}
\noindent$\bullet$ \textbf{Case $\epsilon = 0$ :}

When $\epsilon = 0$ (i.e. for light rays), $U\left(R; n, E, 0\right)$ is always positive and almost constant. So the motion is always possible for all values of $R$ and the light rays might fall in the black hole (if $\dot{R}(0) < 0$), or be diffused (if $\dot{R}(0) > 0$), but no bounded trajectory is possible.
\vspace{4mm}

\noindent$\bullet$ \textbf{Case $\epsilon = 1$ :}

When $\epsilon = 1$ (i.e. for massive particles), one has $U\left(r_h; n, E, 0\right) > 0$ and $U'\left(R; n, E, 0\right) \approx - N'(R) < 0$ then the ``potential'' would be strictly decreasing. Consequently, there would exist one unique $r_\star$ such that $U\left(r_\star; n, E, 0\right) = 0$ and the function would be positive on $\left[r_h, r_\star\right]$ and negative elsewhere. Massive particles would then always be absorbed in this case.

\subsubsection{Case $n = 0$ :}

This case is, in some sense, the most relevant one since one recovers the case of a spherically symmetric space-time for which geodesic motion always occur in a plane which (in an appropriate coordinate system) can be chosen to be the equatorial one.

The function $U\left(R; 0, E, L_3\right)$ is given by
$$U\left(R; 0, E, L_3\right) = \frac{E^2}{A(R)^2} - N(R) \left(\frac{L^2}{R^2} + \epsilon\right).$$
This case is the only one for which motions in the equatorial plane are possible with both $E \neq 0$ and $L_3 \neq 0$.

The analysis of the situations where $E = 0$ or $L_3 = 0$ was already pointed out and was valid for all $n$. We can then focus on the geodesics with $E \neq 0$ \textbf{and} $L_3 \neq 0$.

\vspace{4mm}
\noindent$\bullet$ \textbf{Case $\epsilon = 0$ :}

\begin{figure}[h!]
 \begin{center}
  \includegraphics[scale=0.7]{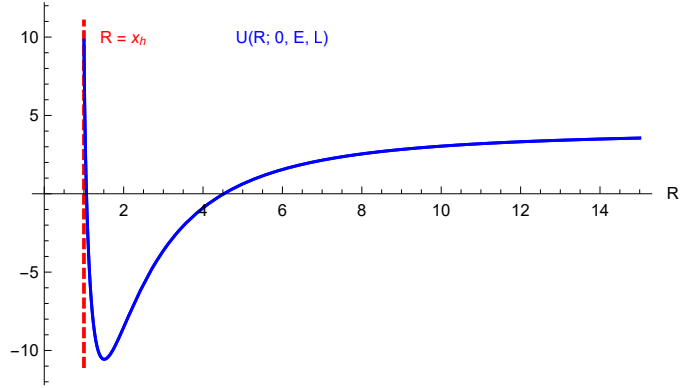}
 \end{center}
 \caption{Shape of the ``potential'' in the exterior space-time for $\epsilon = 0$ and $n = 0$.}
 \label{Uder}
\end{figure}

Fig.~\ref{Uder} shows the shape of $U\left(R; 0, E, L_3\right)$, which is the same for all non-vanishing values of $E$ and $L_3$ when $\epsilon = 0$. 

For a given value of $L_3 = L$, if $E^2$ is sufficiently small (smaller than $V(x_m)$, where $x_m$ is the location of the maximum of $V$\footnote{That is the minimum of $U\left(R; 0, E, L_3\right) \approx E^2 - V(R)$.}), $U\left(R; 0, E, L_3\right)$ possesses two zeros,  $x_1$ and $x_2$ with $x_1 < x_2$, is negative between those values and positive elsewhere, just as in Fig.~\ref{Uder}. Consequently, there are two possible types of motions : if $R(\lambda = 0) < x_1$, the light ray will be absorbed by the black hole (even if $\dot{R}(0) > 0$), and if $R(\lambda = 0) > x_2$, the particle will diffuse (even if $\dot{R}(0) < 0$).

Conversely, if $E^2 \gg V(x_m)$, the curve has the same shape but is always positive. The photon will then be absorbed if $\dot{R}(0) < 0$ and diffused if $\dot{R}(0) > 0$.

Between those situations (when $E^2 \approx V(x_m)$), fine-tuning the energy, there is a limit such that the ``potential'' admits a unique zero equal to its minimum. In this limit, \textit{unstable} circular motions are possible.

\vspace{4mm}
\noindent$\bullet$ \textbf{Case $\epsilon = 1$ :}

One of the distinguished properties of the massive case with respect to the massless case is the
existence of stable circular orbits for large enough values of the angular momentum $L$.
This is a well-known property in the Schwarzschild case where, in our units, stable circular orbits exist for $L^2 \geq 3$. In this case, the smallest stable circular orbit is then reached for $L=\sqrt{3}$, corresponding to $R_c=3$.

We have checked the influence of $\gamma$ on this property. As we already pointed out, the shape of the curve is due to $V(R)$ which depends on $\gamma$, via the metric function $N$, and on $L_3$. Nevertheless, since $\gamma$ does not change the qualitative feature of $N$, its influence on the shape of $V(R)$ would be negligible.

Consequently, since the existence of stable circular orbits and, more generally, of bounded trajectories require the existence of extrema of $V(R)$, existence of such kinds of motions would be controlled by $L_3$ as in the Schwarzschild case and the critical value of $L_3$, say $L_c$, would not vary significantly when the parameter $\gamma$ increases.

\begin{figure}[h!]
 \begin{center}
  \includegraphics[scale=0.7]{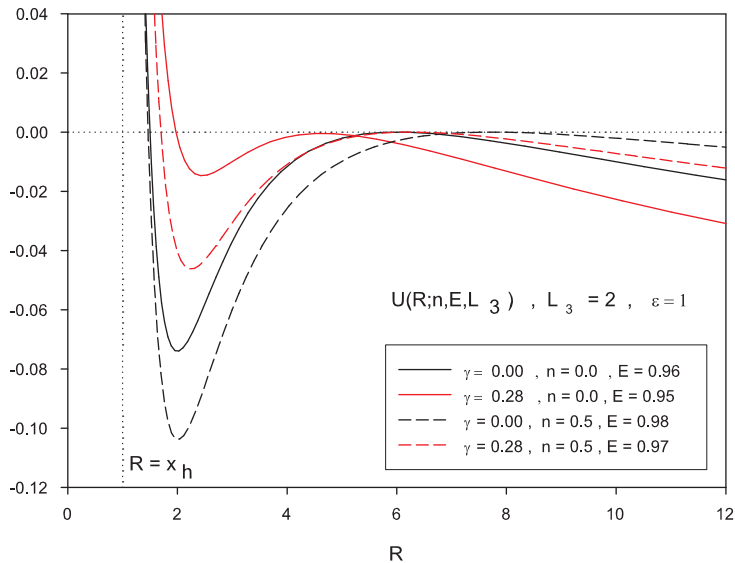}
 \end{center}
 \caption{Shape of the ``potential'' in the exterior space-time for $\epsilon = 1$ and several values of 
  $\gamma$ and $n$. The parameter $E$ was tuned as to coincide with the stable circular orbit. Note : Pay attention that the two dotted curves are unphysical; see discussion in the text.}
 \label{potential}
\end{figure}

Nevertheless, $\gamma$ would have an influence on the position of the extrema of $V(R)$ and the value of $V(R)$ at these extrema. The effective potential in the region of the stable circular orbit is shown on Fig.~\ref{potential} for $L_3 = 2$ and two values of the parameters $n$ and $\gamma$~; we here focus on the solid lines (corresponding to $n=0$). The variation of the potential valley due to the changes of $\gamma$ can be appreciated on the picture.

For a given value of $L \ge L_c$, if we note $x_{\min}$ and $x_{\max}$ the position of the minimum and the maximum of $U(R; 0, E, L_3)$, one has $x_{\min} < x_{\max}$. Our numerical results indicate that, when $\gamma$ increases, $x_{\min}$ increases and $x_{\max}$ decreases while the depth of the potential valley $\left\vert U(x_{\max}; 0, E, L_3) - U(x_{\min}; 0, E, L_3) \right\vert$ also decreases. Consequently, increasing $\gamma$, the range of energy for which bounded trajectories would exist\footnote{The values of $E$ for which $U(x_{\max}; 0, E, L_3) > 0$ and $U(x_{\min}; 0, E, L_3) < 0$.} also decreases. This point is in agreement with our interpretation of section \ref{subsectgenmotion} (see top-left part of Fig.~\ref{vary_gamma} and discussion in the text) : when $\gamma$ is increased ``the black hole becomes more attractive'', since it would be able to absorb particles with significantly higher energy.

\subsubsection{Discussion}

In this paragraph, we have studied particle motions in the equatorial plane. In conclusion to this discussion, let us here summarize and talk about the interpretation of our results.

We saw that, to guarantee a motion confined in the equatorial plane, one have to fulfil condition \eqref{condequator}. Since no motion is possible with $E = 0$, this reduces to
$$n = 0 \lor L_3 = 0.$$

When $n = 0$, the pattern is qualitatively the same as in the Schwarzschild case for all the allowed values of $\gamma$. Increasing $\gamma$ would just quantitatively increase the black hole attraction (see discussion above). In terms of bounded trajectories, massless particles admit only unstable circular orbits, while massive ones admit stable and unstable circular orbits and bounded trajectories, assuming that they have a sufficiently high angular momentum $L_3$.

Our most important result concerns the case $n \neq 0$. In this case, in order to satisfy \eqref{condequator}, one must necessarily have $L_3 = 0$. Then the only possible motions enclosed in the equatorial plane are the purely radial ones. Consequently, there is no possible bounded trajectory in the equatorial plane for $n \neq 0$. This further reinforces the idea that the NUT charge mimic a rotation and produces frame-dragging-like effects.

Actually, when $n \neq 0$, if one wants to obtain a ``potential'' ensuring the existence of stable circular orbit (see dotted curves of Fig.~\ref{potential}), one have to impose $L_3 \ge L_c(n)$. For example, setting $\gamma = 0$, we find that $L_c(n)$ slightly decreases while $n$ increases (e.g. $L_c(n=0) = \sqrt{3}$, $L_c(n=1) \approx \sqrt(2)$), the radius $R_c$ increases and diverges for $n \to 1$. These values do not vary significantly when the parameter $\gamma$ is increased. But, since $L_c(n)$ remains always strictly positive, such a case is not physically possible since it would violate \eqref{condequator}.

\section{Conclusion}

In this paper we have investigated the effects of a NUT-charge on the family of hairy black holes existing in the Einstein-Gauss-Bonnet gravity extended by a real scalar field coupled to the Gauss-Bonnet term. The underlying solutions of the equations form galileon.

In Sect.~\ref{sectnum}, we have seen that the NUT-charge $n$ smoothly deforms the solutions of \cite{Sotiriou:2014pfa}, characterized by the Gauss-Bonnet coupling constant $\gamma$, but affects non-trivially the singularity structure in the interior of the solution. We have put a special attention on the structure of this interior solution and shown that, while it presents two singularities (one located at $x=0$ and another one at $x = x_c(\gamma) > 0$) for any $\gamma \neq 0$ when $n=0$, a non-vanishing NUT-charge tend to regularize the solution for small values of $\gamma$. In particular we have seen that for $n\neq0$, there exists a critical value of the Gauss-Bonnet coupling constant, say $\gamma_{c}(n)$, such that for $\gamma<\gamma_c$ the interior solution presents only one singularity at $x=0$, while for $\gamma>\gamma_c$ a second singularity occurs at a critical radius $x_c>0$. Our numerical results indicate that $\gamma_c$ increases with $n$.

Existence of the critical radius $x_c$ is essential to understand the bound in the domain of existence for the solutions in the $\left(\gamma,n\right)$ plane. Our numerical results tend to proof that for a fixed $\gamma$, when it exists, $x_c$ slightly decreases with $n$ while, for a fixed $n$, it increases with $\gamma$. For a given $n$, this increase of $x_c$ with $\gamma$ was responsible for the existence of a maximal value $\gamma_{\max}(n)$ above which solutions cannot exist. This $\gamma_{\max}$ corresponding to the value of $\gamma$ for which $x_c$ tends to the black-hole horizon radius. The solutions would then stop existing before exhibiting a naked singularity.

Finally, in Sect.~\ref{light-geodesics}, we have characterized the geodesics of massless and massive test particles 
in the space-time of the underlying galileon, finding that, mimicking frame-dragging effects, a non-vanishing NUT-charge gives rise to non-planar geodesics. More than this, we have established that the NUT-charge avoids the existence of motion confined in the equatorial plane. The Gauss-Bonnet parameter haves a quantitative influence on the geodesics but cannot re-establish the properties known in the ``minimal'' Schwarzschild limit in the presence of a NUT-charge.

\pagebreak
\appendix

\section*{Appendix}

This appendix provides several plots to complete the illustrations of the situation described in Sect.~\ref{light-geodesics} for the geodesic motions.

\begin{figure}[h]
 \centering
 \includegraphics[scale=0.27]{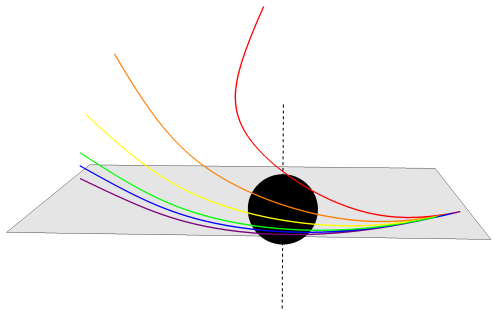} 
 \caption{Geodesics for $\gamma = 0.28$ and for $n=0$, $n=0.05$, $n=0.1$, $n=0.25$, $n=0.5$, $n=1$, respectively represented in purple, blue, green, yellow, orange and red. The setup is identical as in Fig.~\ref{vary_gamma}.}
 \label{vary_n_annexe}
\end{figure}

Fig.~\ref{vary_n_annexe} emphasizes the fact that the particles evolve in a plane if and only if $n = 0$, for all values of $\gamma$.

\begin{figure}[H]
 \begin{minipage}[c]{.3\linewidth}
  \includegraphics[scale=0.3, trim={1cm 1cm 1cm 1cm},clip]{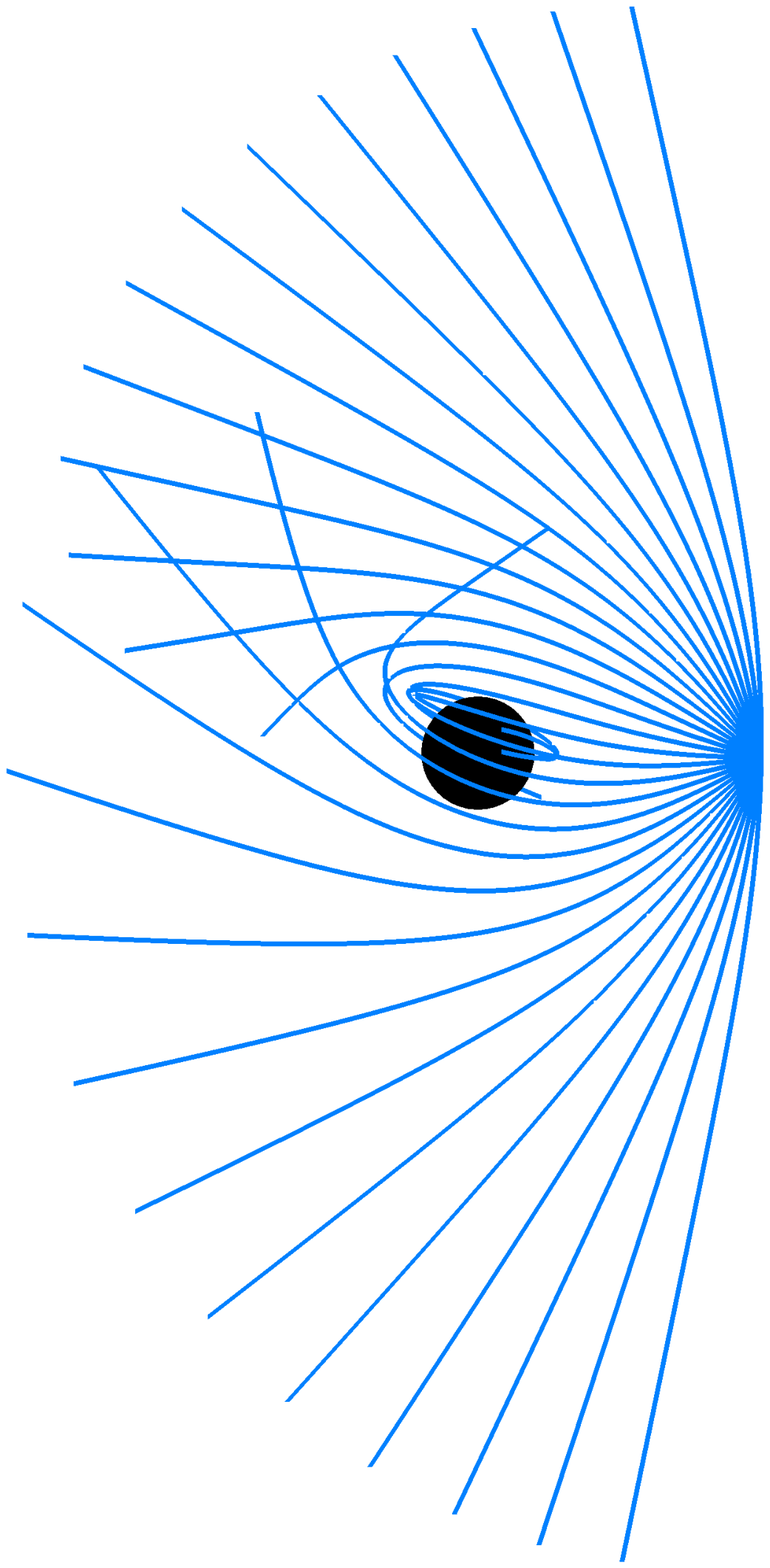}
 \end{minipage} \hfill
 \begin{minipage}[c]{.3\linewidth}
  \hspace{-2cm}\includegraphics[scale=0.3,trim={2cm 0cm 1cm 0cm},clip]{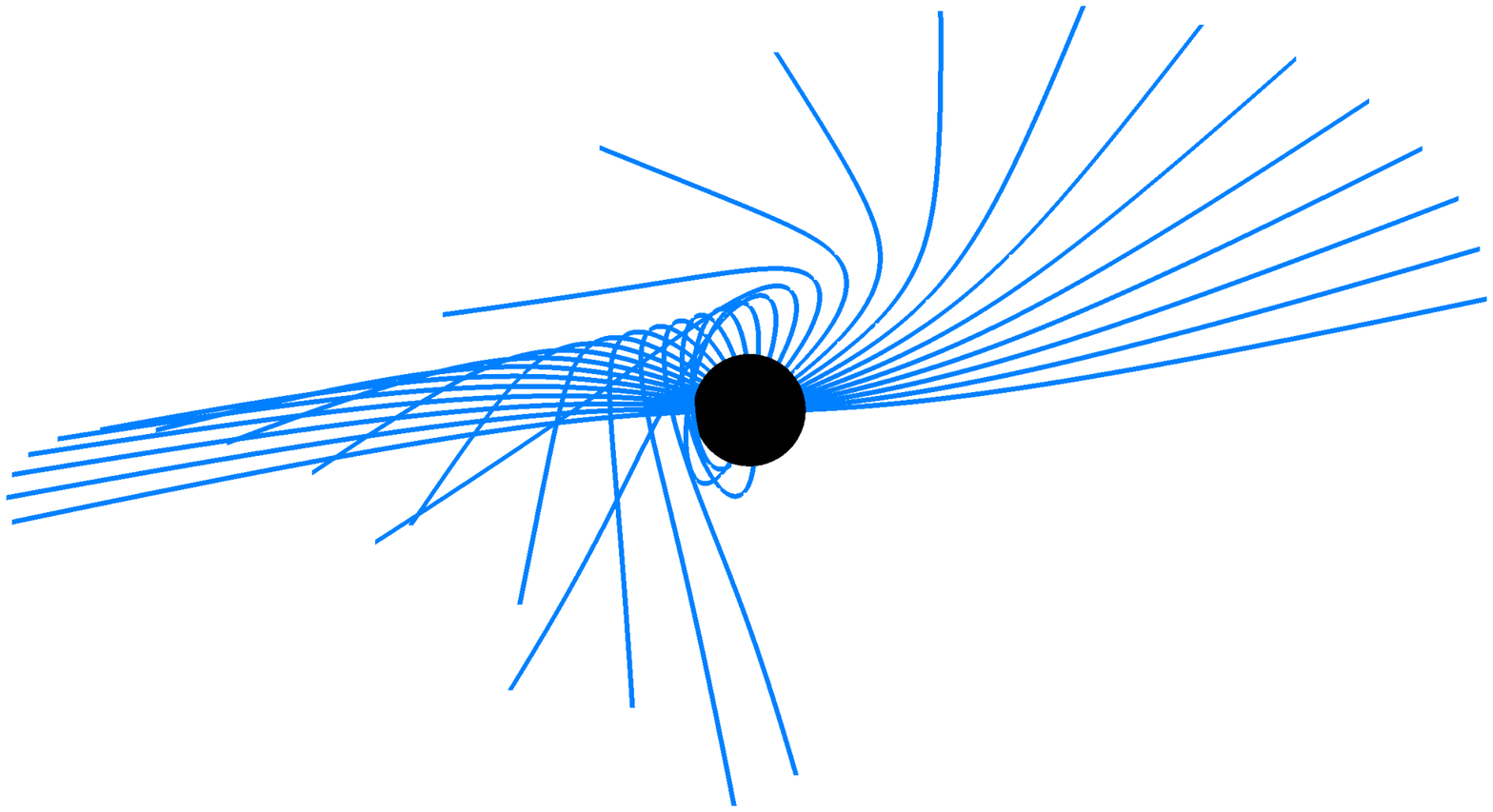}
 \end{minipage}
 \caption{\small{Trajectories for $\gamma = 0.15$ and $n=0.5$. Photons are emitted in the direction $\varphi_s = 0.5$ and for various $\theta_s \in \left[0,\pi\right]$. \textit{Left} and \textit{Right} present the same setup from different points of view.}}
 \label{same_phi}
\end{figure}

On Fig.~\ref{same_phi}, we show various trajectories for a fixed value of $\varphi_s$ and for several values of $\theta_s$, where $\varphi_s$ and $\theta_s$ are respectively polar and azimuthal angle parametrizing the initial direction of the geodesics on the local celestial sphere of the observer. We can clearly see in this plot that when $n \neq 0$ the space-time is not symmetric under the transformation $\theta_s \to \pi - \theta_s$.

\begin{figure}[h]
 \begin{minipage}[c]{.48\linewidth}
  \includegraphics[scale=0.6,trim={1cm 10cm 1cm 10cm},clip]{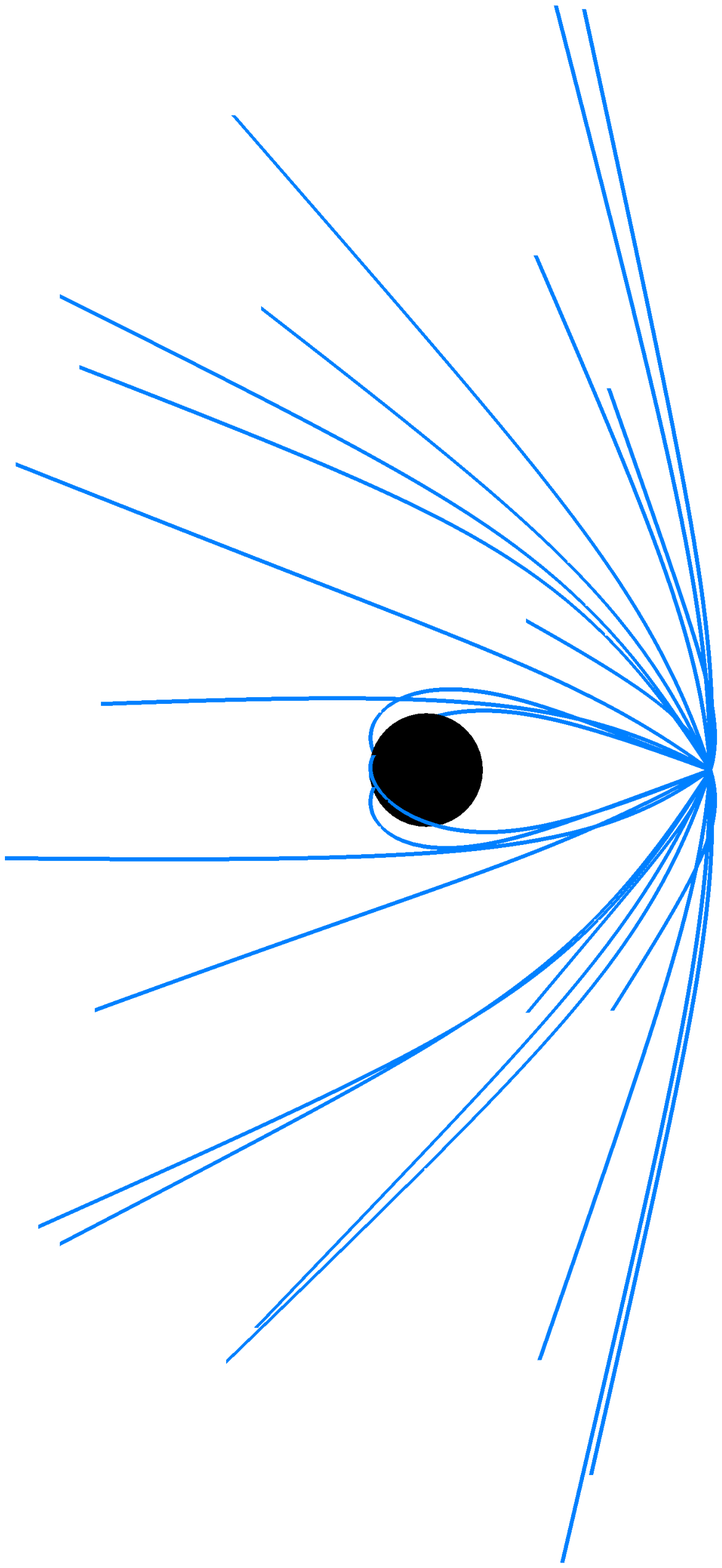}
 \end{minipage} \hfill
 \begin{minipage}[c]{.48\linewidth}
  \includegraphics[width=\textwidth,trim={3cm 3cm 3cm 3cm},clip]{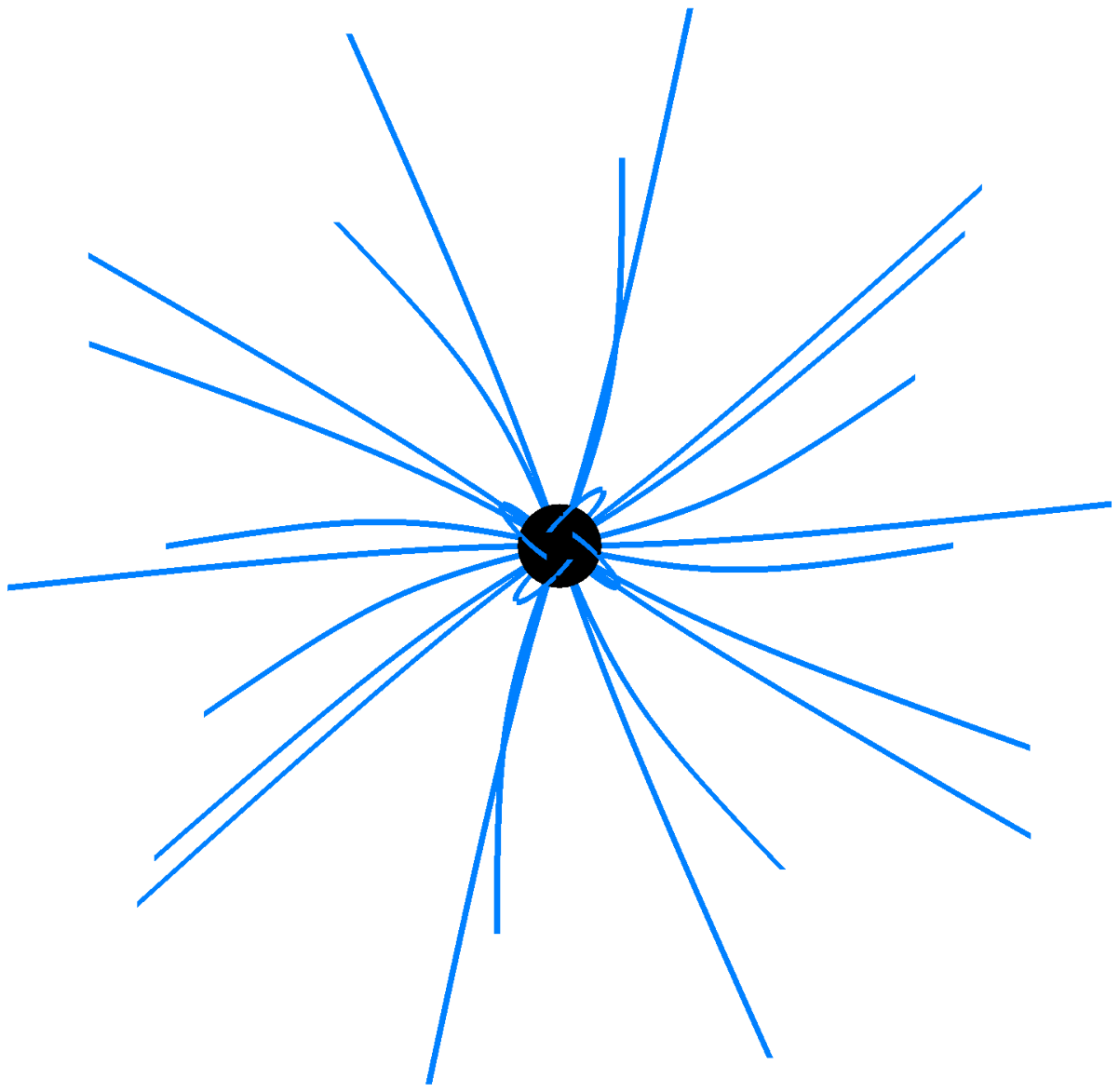}
 \end{minipage}
 \vspace{0.5cm}
 \begin{minipage}[c]{.48\linewidth}
  \includegraphics[scale=0.6,trim={1cm 10cm 1cm 10cm},clip]{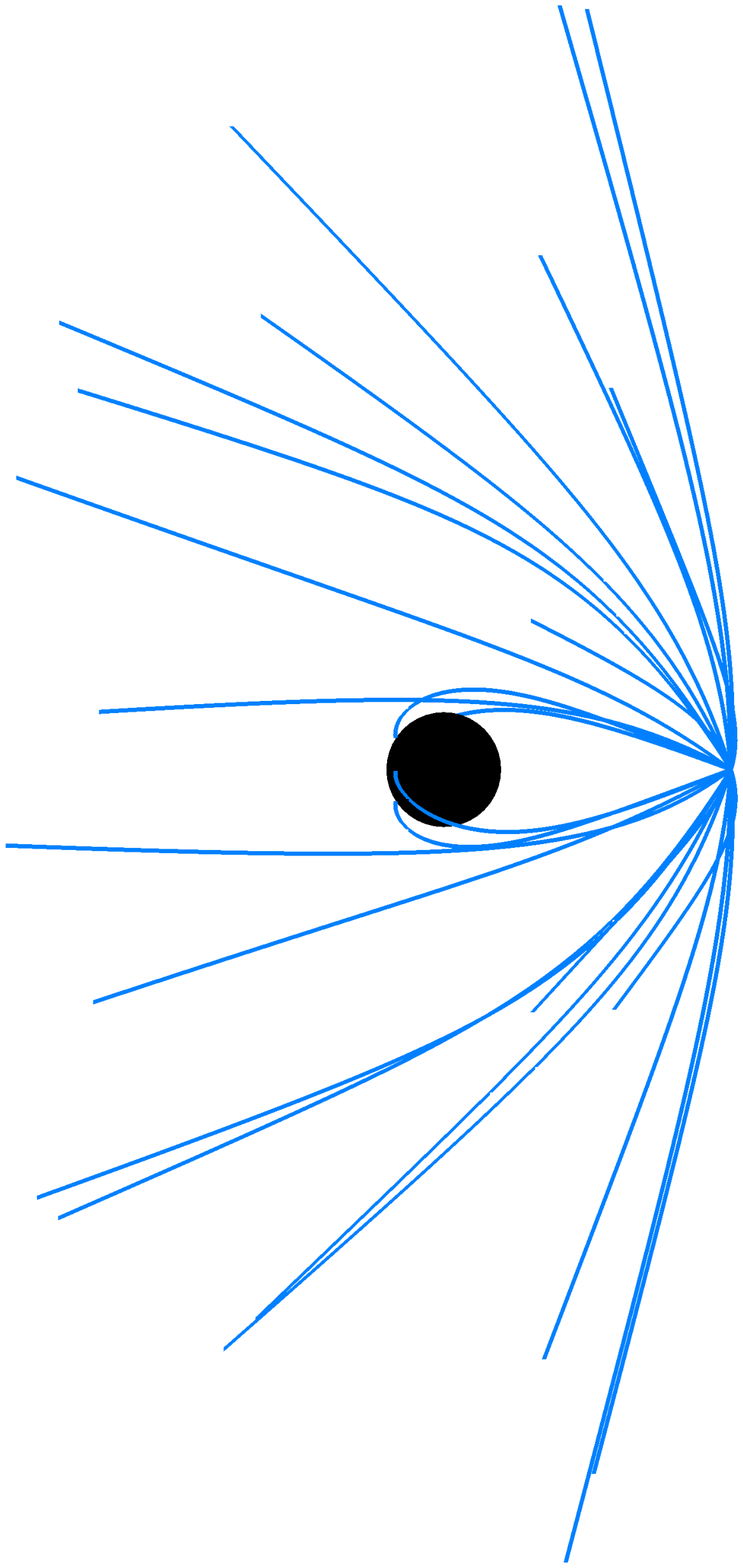}
 \end{minipage} \hfill
 \begin{minipage}[c]{.48\linewidth}
  \includegraphics[width=\textwidth,trim={3cm 3cm 3cm 3cm},clip]{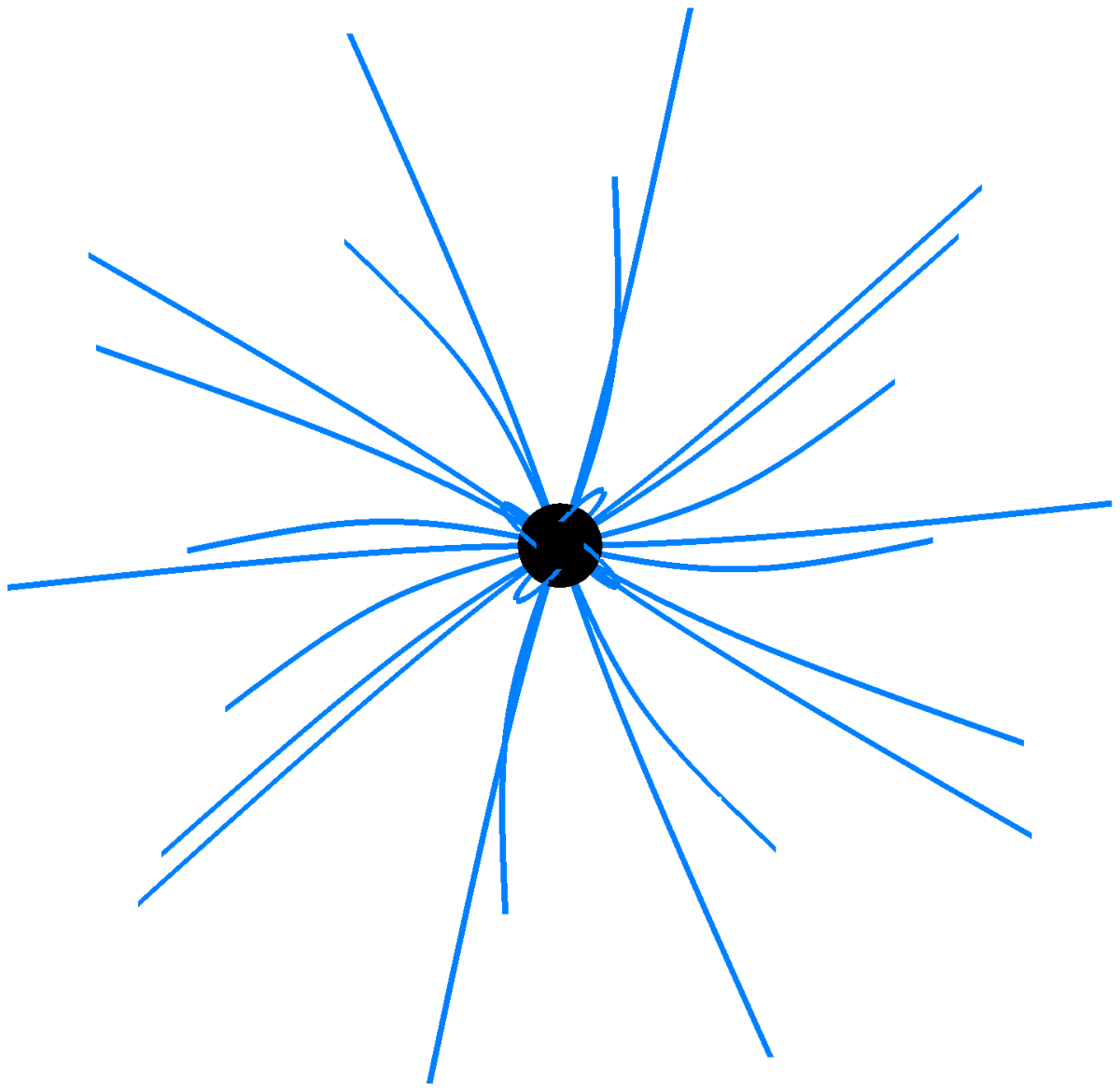}
 \end{minipage}
 \caption{\small{\textit{Top (left and right)} : $\gamma=0.1$. \textit{Bottom (left and right)} : $\gamma=0.28$. Light rays for various initial directions and $n=0.25$ are represented. The observer lies where the trajectories meet. On the top and bottom figures, the left and right situations are the same, but from a different point of view so the behaviour of the geodesics becomes clearer.}}
 \label{spiders_annexe}
\end{figure}

Completing Fig.~\ref{spiders}, Fig.~\ref{spiders_annexe} illustrates the influence of the Gauss-Bonnet parameter on the curvature of light rays for fixed value of the NUT parameter. As in Fig.~\ref{spiders} the purpose of the picture is not to show in detail where go each geodesic. The two cases look qualitatively similar and are analogous to the lower plots in Fig.~\ref{spiders}, confirming that the presence of the NUT charge mimic a rotation. Nevertheless, even if the two plots are qualitatively similar, looking at the right parts, one can see that the frame-dragging effect is influenced by $\gamma$ since the four photons absorbed by the black hole near the centre of the picture are not absorbed at the same spot.

\clearpage

 \end{document}